\documentclass[12pt,a4paper]{article}
\pdfoutput=1
\usepackage{enumitem}

\usepackage{amsmath,amscd}
\usepackage{amsfonts}
\usepackage{amssymb,scalerel,stackengine}
\usepackage{amsmath,amsfonts,amssymb,amsthm,bbm,bm}
\usepackage{relsize}
\usepackage{empheq}
\usepackage{graphicx,shortvrb}
\usepackage{comment}
\usepackage{cite}
\usepackage[utf8]{inputenc}
\usepackage[colorlinks=true,linkcolor=blue, linktoc=page, urlcolor=blue,citecolor=magenta]{hyperref}

\usepackage{xcolor}

\numberwithin{equation}{section}

\hypersetup{
    colorlinks,%
    citecolor=blue,%
    filecolor=blue,%
    linkcolor=blue,%
   urlcolor=blue,
   linktoc=page
}

\newcommand\widecheck[1]{%
\savestack{\tmpbox}{\stretchto{%
  \scaleto{%
    \scalerel*[\widthof{\ensuremath{#1}}]{\kern-.6pt\bigwedge\kern-.6pt}%
    {\rule[-\textheight/2]{1ex}{\textheight}}
  }{\textheight}%
}{0.5ex}}%
\stackon[1pt]{#1}{\scalebox{-1}{\tmpbox}}%
}

\newcommand{\bea}{\begin{eqnarray}}
\newcommand{\eea}{\end{eqnarray}}

\definecolor{darkgreen}{rgb}{0.0, 0.8, 0.1}

\newcommand{\spa}{\ , \ \ }

\begin{document}

\setcounter{tocdepth}{2}

\begin{titlepage}

\begin{flushright}\vspace{-3cm}
{\small
\today }\end{flushright}
\vspace{0.5cm}

\begin{center}

\begin{center}
{\Large{\bf{Moving away from the Near-Horizon Attractor \\[1mm] of the Extreme Kerr Force-Free Magnetosphere }}}
\end{center}

\begin{center}
\line(1,0){450}
\end{center}

\centerline{\bf{
F.~Camilloni$^{\dagger\,\ddagger}$\footnote{email: filippo.camilloni@nbi.ku.dk},
G.~Grignani$^\dagger$\footnote{email: gianluca.grignani@unipg.it},
T.~Harmark$^\ddagger$\footnote{email: harmark@nbi.ku.dk},
R.~Oliveri$^*$\footnote{email: roliveri@fzu.cz},
M.~Orselli$^{\dagger\,\ddagger}$\footnote{email: orselli@nbi.dk}
}}\vspace{2pt}

\vspace{2mm}
\normalsize
\bigskip\medskip
\textit{{}$^\dagger$ Dipartimento di Fisica e Geologia, Universit\`a di Perugia, I.N.F.N. Sezione di Perugia, \\ Via Pascoli, I-06123 Perugia, Italy }\\ \medskip
\textit{{}$^\ddagger$ Niels Bohr Institute, Copenhagen University,\\  Blegdamsvej 17, DK-2100 Copenhagen \O{}, Denmark}\\ \medskip
\textit{{}$^*$ CEICO, Institute of Physics of the Czech Academy of Sciences,\\
Na Slovance 2, 182 21 Praha 8, Czech Republic }

\vspace{10mm}

\begin{abstract}
\noindent
{We consider force-free magnetospheres around the extreme Kerr black hole. 
In this case there is no known exact analytic solution to force free electrodynamics which is stationary, axisymmetric and magnetically-dominated.
However, any stationary, axisymmetric and regular force-free magnetosphere in extreme Kerr black hole approaches the same attractor solution in the near-horizon extreme Kerr (NHEK) limit with null electromagnetic field. We show that by moving away from the attractor solution in the NHEK region, one finds magnetically-dominated solutions in the extreme Kerr black hole with finite angular momentum outflow. This result is achieved using a perturbative analysis up to the second order.}
\end{abstract}

\end{center}

\end{titlepage}

\tableofcontents

\setcounter{footnote}{0}

\section{Introduction and outline}
\label{Sec:Intro}
%
It is by now well established that many astrophysical objects emitting highly energetic collimated jets, such as active galactic nuclei and pulsars, must contain at their center a compact rotating source, a black hole or a magnetized neutron star. 
The magnetosphere of these objects is filled with plasma, but, in general, the plasma energy density can be neglected compared to the energy density of the electromagnetic field. Consequently, one can ignore the Lorentz four-force density, the current can be traded for the covariant derivative of the electromagnetic field strength $F$ through Maxwell's equations and the resulting dynamics of the electromagnetic field is governed by the equations of force free electrodynamics (FFE); for a review see, \emph{e.g.},~\cite{Beskin:2010iba,Gralla:2014yja} and references therein. This framework provides the minimal nontrivial level of description for pulsar \cite{Goldreich} and black hole magnetospheres \cite{Blandford:1977ds} in which the plasma is assumed to be in equilibrium with a magnetically-dominated electromagnetic field ($B^2>E^2$, \emph{i.e.}, $F^2>0$). The FFE approximation is an effective description of the magnetosphere in the funnel region around jets, and it is supported by magneto-hydrodynamics (MHD) numerical simulations \cite{McKinney_2004,KomissarovMHD,McKinney_2005,KomissarovMHD2,McKinney_2007,Tchekhovskoy_2008,Tchekhovskoy:2009ba,Tchekhovskoy2011}.
Nevertheless, FFE equations are still too complicated to be solved analytically in a Kerr black hole background. In a suitable gauge, consistent with the stationarity and axisymmetry of the Kerr metric, the FFE equations can be expressed in terms of three functions that are related to some of the components of the electromagnetic field strength: the so called stream function $\psi(r,\theta) \equiv A_\phi(r, \theta) $, that represents the magnetic flux through a loop surrounding the rotation axis defined by $(r,\theta)$, the angular velocity of the field lines $\Omega(\psi)$, which is a function of $\psi$, and $I(\psi)$, that represents the poloidal electric current, defined as the electric current flowing upwards through the loop around the rotation axis, which is also a function only of $\psi$.
There are several classes of known analytic solutions to FFE in flat spacetime \cite{1973ApJ...180..207M} and in the Schwarzschild black hole background~\cite{Lyutikov:2011tq} for the stream function $\psi$ with suitable choices of $\Omega(\psi)$ and $I(\psi)$. In contrast, in the case of the Kerr black hole metric, the only known classes of analytic solutions to FFE are those of Dermer and Menon~\cite{Menon:2005va,MD2005,Menon_2011,Menon:2015dea} and of Brennan, Gralla and Jacobson~\cite{BGJ}. These have the common property of being null, so that, not only $\vec{E}\cdot\vec{B}=0$, which is implied by the FFE equations, but also $E^2=B^2$, namely $F^2=0$. Physically acceptable solutions, however, should be magnetically-dominated ($F^2>0$) in order for the equations of motion to be hyperbolic \cite{Komissarov,Palenzuela_2011}. 
The magnetosphere and the accretion disc surrounding black holes are believed to be responsible for the production of the jet, which can be very energetic for a highly spinning black hole \cite{Tchekhovskoy2011,McKinney_2012,Penna_2013}. The mechanism that is thought to be responsible for the jet production is the Blandford and Znajek (BZ) process~\cite{Blandford:1977ds}, in which the rotational energy of a black hole immersed in a magnetic field, supported by the accretion disc, is converted into the energy that feeds the jet \cite{Komissarov_2004}. Black holes immersed in magnetic fields could have a force-free (FF) plasma and the presence of such a plasma enables an electromagnetic Penrose process in which even stationary fields can efficiently extract energy, especially from a highly spinning black hole. Therefore, for astrophysical purposes, it is very important to study FFE in the background of a maximally-rotating (extreme), or nearly extreme,  Kerr black hole.  
Moreover, the BZ process operates close to the horizon and it is localized at that physical scale. BZ actually found an approximate analytic solution in the opposite regime, in a perturbative expansion valid at small spin.
For recent attempts to extend the BZ analytic perturbative analysis to higher orders in the rotation parameter, see \cite{Tanabe:2008wm,Pan:2015haa,Pan:2015iaa,Grignani:2018ntq,Armas:2020mio}.
The finite spin version has been studied analytically in \cite{GHO2019} and numerically for example in \cite{Komissarov:2001sjq,Tchekhovskoy:2009ba,Tchekhovskoy2011,Nathanail:2014aua}. 
Our goal with this paper is to implement the new strategy proposed in our letter \cite{CGHOOletter} for finding magnetically-dominated FF magnetospheres in the background of extreme Kerr, in which $J=M^2$, where $J$ is the angular momentum and $M$ is the mass of the black hole.
This choice is motivated by the observational evidence that nearly extreme black holes exist in nature \cite{McClintock_2006,Gou_2011,Brenneman_2013,Gou_2014,Reynolds_2013}.

The near horizon region of extreme Kerr geometry (NHEK) results to be endowed with an enhanced symmetry group as compared to the generic Kerr metric: in the NHEK limit, the time symmetry of the Kerr black hole is enhanced to a global $SO(2,1)$ conformal symmetry~\cite{Bardeen:1999px}. Such a global symmetry led to the discovery of several infinite families of FFE analytic solutions \cite{Lupsasca:2014pfa,Lupsasca:2014hua,Zhang_2014,Compere:2015pja,Oliveri:2018dfo}.
Among the infinite solutions of FFE in the NHEK limit~\cite{Compere:2015pja} there is one, first found in \cite{Lupsasca:2014hua}, that singles out as particularly interesting. 
Starting from the general form of a stationary axisymmetric Maxwell field strength $F$ on extreme Kerr background, sourced by a current, and which is regular on the future event horizon, one ends up with a field that is highly constrained in the NHEK limit~\cite{Gralla:2016jfc}. The resulting field strength, that in our letter \cite{CGHOOletter} we dubbed as ``attractor solution", is still stationary and axisymmetric, but it is also null ($F^2=0$) and contains an arbitrary function that depends only on the polar angle $\theta$. This arbitrary function can then be fixed in terms of the poloidal current by requiring that the solution is FF or, equivalently, by imposing the so called Znajek condition at the horizon~\cite{10.1093/mnras/179.3.457}. Therefore, among the infinite known analytic FFE solutions in the NHEK background, the attractor solution is the only one that can be actually connected to a possible stationary axisymmetric solution on the extreme Kerr black hole. 
Due to the non-linearity of the FFE equations in a Kerr background, it is extremely difficult to construct an analytic magnetically-dominated ($F^2>0$) solution. For this reason, in this paper, we will take advantage of the attractor solution to construct perturbative solutions to FFE around it. As proposed in our letter \cite{CGHOOletter}, the strategy that we follow is to move away from the throat in the NHEK geometry towards extreme Kerr, with the aim of finding, at the second order in a suitable expansion parameter, a perturbative magnetically-dominated solution to FFE in the extreme Kerr background. 
A relevant result that we obtain in the journey from NHEK towards extreme Kerr, is that the Lorentz invariant $F^2$ can be positive; the perturbative corrections in fact give rise to a Maxwell field strength $F$ that is magnetically-dominated. We will show this explicitly by performing computations up to second order in our perturbative expansion.
In particular, we will find that the FFE equations lead to a differential equation for the second post-NHEK order correction to the stream function $\psi_2(\theta)$ which, with some suitable ansatz, can be solved exactly.
The solution that we obtain in this way contains some parameters that can then be fixed to render the solution magnetically-dominated. This is the main result of this paper: we have shown that, starting with a null solution to FFE in the NHEK geometry, one can construct perturbative solutions to FFE in the extreme Kerr background which are magnetically-dominated with finite angular momentum outflow.
The ansatz we used, however, even though it allows us to analytically solve the FFE equations up to the second post-NHEK order and to compute $F$ and $F^2$ explicitly, has the disadvantage of making $F$ not regular on the rotation axis even if $F^2$ is regular and can be made positive. The regularity issue of the field strength at the rotation axis might be resolved by a different ansatz and/or by solving the boundary-value problem by taking into account the presence of the inner light-surface.
The derivation of the differential equation for $\psi_2$ is by itself a result. It is a well defined differential equation in $\theta$ for which we looked for analytic solutions, but it could be also studied numerically, with more physical boundary conditions that give a field strength $F$ regular on the rotation axis. This is beyond the scope of the present paper, but it would certainly be an interesting project for the future.
The paper is organized as follows. In Sec.~\ref{section:FFE}, we review the Kerr black hole background and its near-horizon geometry. 
We then briefly discuss FFE and comment on the Znajek condition that ensures the regularity of the field at the event horizon. 
In Sec.~\ref{Sec:Attractor}, we discuss the ``attractor mechanism'' described in the introduction, we present the attractor solution and the expansion of the field variables $\psi$, $\Omega$ and $I$ around it. The NHEK solutions at the zeroth order in the expansions are presented here. In Sec.~\ref{Sec:Climbing}, we derive the post-NHEK corrections, we solve exactly the first order corrections and we present the second order solutions for $I$ and $\Omega$ and the differential equation for $\psi_2$.
Sec.~\ref{SubSec:MD} shows how the Menon and Dermer type of solutions corresponds to neglecting radial contributions in our perturbative scheme. In Sec.~\ref{SubSec:Ansatz 1}, we present new perturbative solutions. We introduce ansatzes that allows us to solve exactly the differential equation for $\psi_2$ and this in turn leads to the calculation of $F^2$. It is then shown that $F^2$ can be positive for a certain  choice of the parameters, so that the corresponding perturbative solution becomes magnetically-dominated with finite angular momentum extraction. 
Finally, we conclude with a summary of our results in Sec.~\ref{Sec:Conclusion}. In Appendix~\ref{App:A}, we present the explicit expressions for the fields and constraints in the perturbative expansion. In Appendix~\ref{App:B}, we discuss the zeroth and the first post NHEK orders in the case of $\psi_0={\rm{const}}$. Appendix \ref{Expr2ndNHEK} contains lengthy expressions entering the second post-NHEK order computation.
\paragraph*{Notation:} 
We fix geometric units such that $G=1=c$.
We adopt the notation that a quantity, say $Q(r, \theta; a)$, when evaluated at the event horizon $r_+ = r(a)$ is denoted by $Q_+ \equiv Q(r_+,\theta; a)$, while when evaluated at the event horizon and at extremality is denoted by $Q_0 \equiv Q(M,\theta; M)$.
%


\section{Force-free electrodynamics around Kerr black holes} \label{section:FFE}

In this section, we review briefly force-free electrodynamics (FFE) around Kerr black holes (see, \emph{e.g.}, \cite{Gralla:2014yja} and references therein) and the near-horizon extreme Kerr (NHEK) geometry \cite{Bardeen:1999px}. 

\subsection{Kerr and NHEK geometry} \label{Sec:Geometry}

The metric for a Kerr black hole with mass $M$ and angular momentum $J$ in Boyer-Lindquist (BL) coordinates is
\begin{align} \label{KerrBL}
ds^2&=-\Big(1-\frac{r_0r}{\Sigma}\Big)dt^2-\frac{2r_0r}{\Sigma}a\sin^2 \theta ~dt d\phi +\frac{(r^2+a^2)^2-a^2\Delta\sin^2 \theta}{\Sigma}\sin^2 \theta ~d\phi^2\nonumber\\
&\quad +\frac{\Sigma}{\Delta}dr^2+\Sigma d\theta^2,
\end{align}
with $r_0 = 2M$, $a=J/M$, and
\begin{equation}
\Sigma = r^2+a^2 \cos^2 \theta \spa  \Delta=(r-r_+)(r-r_-) \, \spa r_\pm = \frac{r_0}{2}\left(1\pm \sqrt{1-\frac{4a^2}{r_0^2}}\right).
\end{equation}
The angular momentum of the black hole is bounded by its mass from the requirement that $a^2 \leq r_0^2/4$. 
When this bound is saturated, we obtain the extreme Kerr black hole with the maximal angular momentum $|J|=M^2$.
In this case, the two horizons coincide at $r_+ = r_-=r_0/2$ and the angular velocity of the black hole, $\bar{\Omega}_+=a/(r_+^2+a^2)$, reduces to $\bar{\Omega}_0=r_0^{-1}$. 
The Kerr black hole spacetime is stationary and axisymmetric corresponding to the two commuting Killing vector fields $\partial_t$ and $\partial_{\phi}$. These Killing vectors span a surface that we refer to as the toroidal surface, while the orthogonal surface to the Killing vectors, spanned by $(r, \theta)$ coordinates, is referred to as the poloidal surface. 
Therefore, the Kerr metric admits the following decomposition into poloidal and toroidal metrics
\begin{equation}
ds^2=ds^2_T+ds^2_P, \quad ds^2_T=g_{tt} dt^2+2g_{t\phi}dtd\phi+g_{\phi\phi}d\phi^2,\quad ds^2_P=g_{rr}dr^2+g_{\theta\theta}d\theta^2,
\end{equation}
which we will make use of in Sec.~\ref{Sec:FFE} for stationary and axisymmetric magnetospheres.
For future reference, the determinant of the metric is the product of the single determinants $g=g_T\cdot g_P$ and, explicitly, in BL coordinates we have
\begin{equation}
g=-\Sigma^2\sin^2\theta,\quad
g_T=g_{tt}g_{\phi\phi}-(g_{t\phi})^2=-\Delta\sin^2\theta,\quad g_P=g_{rr}g_{\theta\theta}=\frac{\Sigma^2}{\Delta}.
\end{equation}
In this paper, we focus on the region near the horizon of an extreme Kerr black hole corresponding to the NHEK geometry \cite{Bardeen:1999px}.
To derive the NHEK geometry, one has to zoom in close to the horizon while corotating with its angular velocity. To this end, one defines first the corotating coordinates
\begin{equation}
t'=\bar{\Omega}_+ t,\quad r'=\frac{r-r_+}{r_+},\quad\Phi=\phi-\bar{\Omega}_+ t \,.
\end{equation}
Then, upon imposing the extreme condition $a=M$ (and thus $\bar{\Omega}_0 = r_0^{-1}=(2M)^{-1}$), one defines the scaling coordinates $(T,R,\theta,\Phi)$ as 
\begin{equation} \label{scaling}
T=\lambda t' = \lambda \frac{t}{r_0},\quad R=\frac{1}{\lambda}r' = \frac{2 r-r_0}{\lambda r_0},\quad\Phi=\phi-\frac{t}{r_0}.
\end{equation}
The NHEK geometry is then achieved by taking the $\lambda\to 0$ limit while keeping the coordinates $T$, $R$, $\theta$ and $\Phi$ fixed. The resulting NHEK metric reads
\begin{equation}\label{NHEK}
ds^2=\frac{r_0^2}{2}\Gamma(\theta)\bigg[-R^2 dT^2+\frac{dR^2}{R^2}+d\theta^2+\Lambda^2(\theta)\big(d\Phi+RdT\big)^2\bigg],
\end{equation}
where we introduced the following functions
\begin{equation}
\label{GammaLambda}
\Gamma(\theta)=\frac{1+\cos^2\theta}{2},\quad \Lambda(\theta)=\frac{2\sin\theta}{1+\cos^2\theta}.
\end{equation}
The event horizon is located at $R=0$.
An important property of the NHEK geometry is that its isometry group is enhanced from the Kerr isometry group $\mathbb{R} \times U(1)$ to $SO(2,1) \times U(1)$ \cite{Bardeen:1999px,Kunduri:2007vf}.
For further details about the NHEK geometry, we refer the reader to \cite{Guica:2008mu,Bredberg:2011hp,Compere:2012jk}.


\subsection{Force-free electrodynamics}
\label{Sec:FFE}

The equations of FFE are the Maxwell equations
\begin{equation}\label{Maxwell}
D_\mu F^{\mu\nu}=j^\nu, \quad D_{[\rho}F_{\mu\nu]} = 0,
\end{equation}
supplemented with the force-free (FF) constraint
\begin{equation}\label{FFE0}
F_{\mu\nu}j^\nu=0.
\end{equation}
Here, $F_{\mu\nu}$ is the electromagnetic field strength, $F_{\mu\nu}=2\partial_{[\mu}A_{\nu]}$ with $A_\mu$ being the gauge potential, and $j^\mu$ the current which we assume is different from zero.
We assume that the electromagnetic field is stationary and axisymmetric around the same rotation axis as the Kerr black hole. This means that we can choose a gauge where
$\partial_t A_\mu=0=\partial_\phi A_\mu$. We define the \emph{magnetic flux} $\psi(r,\theta)$ and the \emph{poloidal current} $I(r,\theta)$ as
\begin{equation}\label{defpsi}
\psi=A_\phi  \spa I=\sqrt{-g}F^{r \theta}.
\end{equation}
From Eq.~\eqref{FFE0}, and using the inhomogeneous Maxwell equations, we notice that the FF constraint is nonlinear and given by
\begin{equation}\label{FFE}
F_{\mu\nu}D_\rho F^{\nu\rho}=0.
\end{equation}
Combining the toroidal $\mu=t,\phi$ components of Eq.~\eqref{FFE}, we get the condition $\partial_rA_t\partial_\theta\psi=\partial_\theta A_t\partial_r\psi$, implying that $A_t$ is a function of $\psi$. We thus define the \emph{angular velocity of the magnetic field lines} $\Omega(r,\theta)$ as $\partial_rA_t=-\Omega \partial_r\psi, \partial_\theta A_t=-\Omega \partial_\theta\psi$,
from which one can infer the integrability condition
\begin{equation} \label{OmegaIntCond}
\partial_r\Omega\partial_\theta\psi=\partial_\theta\Omega\partial_r\psi,
\end{equation}
\emph{i.e.}, $\Omega$ is a function of $\psi$. This latter requirement is equivalent to the Bianchi identity for the 2-form $F$.
From the $\mu=r,\theta$ component of Eq.~\eqref{FFE}, one gets the integrability condition 
\begin{equation} \label{CurrentIntCond}
\partial_rI\partial_\theta\psi=\partial_\theta I\partial_r\psi,
\end{equation}
which implies that also $I$ is a function of $\psi$.
		
It is possible to show that one can always recast a stationary and axisymmetric FF field strength in the form~\cite{Gralla:2014yja}
\begin{equation}\label{FKerrgen}
F= I(\psi) \sqrt{-\frac{g_P}{g_T}} dr\wedge d\theta+d\psi\wedge\big(d\phi-\Omega(\psi)dt\big),
\end{equation}
where the field variables $\psi(r,\theta)$, $\Omega(\psi)$ and $I(\psi)$ are related to each other through the so-called \emph{stream equation} 
\begin{equation}\label{StreamEq}
\Big[\partial_\rho(\sqrt{-g}F^{\phi\rho})-\Omega\partial_\rho(\sqrt{-g}F^{t\rho})\Big]+F_{\theta r}\frac{d I}{d\psi}=0.
\end{equation}
Hereafter, we will consider the field \eqref{FKerrgen} in Kerr spacetime. It explicitly reads as 
\begin{equation}\label{FKerr}
F=\frac{\Sigma I(\psi)}{\Delta\sin\theta}dr\wedge d\theta+d\psi\wedge\big(d\phi-\Omega(\psi)dt\big).
\end{equation}
For sake of completeness, we also report the expression for the invariant $F^2$ in Kerr spacetime
\begin{equation}
    F^2=\frac{2I^2(\psi)}{\Delta\sin^2\theta}+\frac{2\Delta\Sigma^2-2\sin^2\theta \left[a r r_0-((a^2+r^2)^2-a^2\Delta\sin^2\theta)\Omega(\psi)\right]^2}{\Delta\Sigma^2\sin^2\theta\left[(a^2+r^2)^2-a^2\Delta\sin^2\theta\right]}\left[(\partial_\theta\psi)^2+\Delta(\partial_r\psi)^2\right].
\end{equation}
Physically acceptable solutions must have $F^2>0$. In this case, the field strength $F$ is said to be magnetically-dominated. Otherwise, it is said to be electrically-dominated if $F^2<0$ or null if $F^2=0$. Clearly, $F^2$ is a function of $(r, \theta)$. While the first term is always positive or null outside the event horizon, the second term can be anything. Thus, finding solutions to the non-linear stream equation \eqref{StreamEq} such that $F^2$ is positive is a hard task.

Finally, we also mention that the inflow of energy and angular momentum across the event horizon read as \cite{Gralla:2014yja}
\begin{subequations}\label{EJextraction}
\begin{align}
    \frac{dE}{dt} &= 2\pi \int_{0}^{\pi} \Omega_+(\psi)\left(\bar{\Omega}_+ - \Omega_+(\psi) \right)\left(\partial_{\theta}\psi_+ \right)^2\sqrt{\frac{g_{\phi\phi}}{g_{\theta\theta}}}\Bigg|_{r=r_+} d\theta,\\
    \frac{dJ}{dt} &= 2\pi \int_{0}^{\pi} \left(\bar{\Omega}_+ - \Omega_+(\psi) \right)\left(\partial_{\theta}\psi_+ \right)^2\sqrt{\frac{g_{\phi\phi}}{g_{\theta\theta}}}\Bigg|_{r=r_+} d\theta.
\end{align}
\end{subequations}
These expressions account for the energy and angular momentum extraction from the black hole (negative inflow across the horizon) by means of the BZ process.


\subsection{Comment on the Znajek condition at extremality} \label{SubSec:ZC}

We are interested in studying stationary axisymmetric FF fields in the NHEK geometry. An important condition that one has to take into account is the so-called Znajek condition \cite{10.1093/mnras/179.3.457}, which imposes regularity of the electromagnetic field at the event horizon.

For any stationary axisymmetric FF field in Kerr background, the Znajek condition relates $\psi$, $\Omega$ and $I$ on the  event horizon in the following way
\begin{equation}\label{ZC}
\left(I\Sigma-\Lambda \Gamma (r^2 + a^2)(\Omega -\bar{\Omega})\partial_\theta\psi\right)\Big|_{r=r_+}=0.
\end{equation}

In the extreme case, the event horizon is located at $r_0/2$ and the Znajek condition \eqref{ZC} becomes
\begin{equation}\label{ZCExt}
I_0=\frac{\Lambda}{r_0}\left(r_0\Omega_0-1\right)(\partial_\theta\psi)_0,
\end{equation}
where $I_0$, $\Omega_0$ and $(\partial_\theta\psi)_0$, respectively, refer to $I$, $\Omega$ and $\partial_\theta\psi$ evaluated at the event horizon. 
Furthermore, in the extreme case, the Znajek condition must be supplemented with a second necessary condition to ensure the regularity of the field at the event horizon (see Eq.~(120) in \cite{Gralla:2014yja} for details) which is
\begin{equation}\label{ZCExt2}
 (\partial_r I)_0= \frac{\Lambda}{r_0}\left[\left(r_0\Omega_0-1\right)(\partial_r\partial_\theta\psi)_0 + \left(r_0(\partial_r\Omega)_0 - \Lambda^2\Gamma\Omega_0 + \frac{2}{ r_0 \Gamma}\right)(\partial_\theta\psi)_0\right].
\end{equation}
%


\section{The NHEK attractor solution}
\label{Sec:Attractor}

In this section we zoom into the NHEK region of any given stationary, axisymmetric and regular FF magnetosphere \eqref{FKerr} around extreme Kerr. In doing this, one ends up always with the same null and self-similar FF solution in the NHEK geometry, that we dubbed the {\sl attractor solution} \cite{CGHOOletter}. This happens irrespectively of whether one starts in extreme Kerr with a magnetically-dominated solution or not. This result falls into the general argument, presented in \cite{Gralla:2016jfc}, where it has been shown that the limiting field must be stationary, axisymmetric, null and self-similar.
The attractor solution will be our starting point for moving away from the NHEK geometry.

We consider a field strength, $F$, which is stationary, axisymmetric and regular on the future event horizon in the extreme Kerr geometry \eqref{FKerr}. Its behavior near the horizon can be determined by making use of the scaling coordinates \eqref{scaling} and expanding for small $\lambda$. The field is formally expanded as \cite{Gralla:2016jfc}
\begin{align}
F &=\sum_{n=-1}^{\infty}\lambda^{n}F^{(n)}.
\end{align}
The leading-order term $F^{(-1)}$ represents the field in the NHEK geometry \eqref{NHEK} and it is explicitly given by
\begin{equation} \label{Fm1}
F^{(-1)}=\left[\frac{r_0 I_0}{\Lambda} \frac{dR}{R^2} +\big(r_0\Omega_0-1\big)\psi'_0 ~dT \right] \wedge d\theta.
\end{equation}
Here, and in the following, ``prime" denotes derivative with respect to $\theta$.
In deriving Eq.~\eqref{Fm1}, we assumed that $\psi$, $I(\psi)$, and $\Omega(\psi)$ admit a regular expansion near the horizon which, in the case of extreme Kerr, is of the form
\begin{subequations} \label{expansions}
\begin{align} 
\label{psiexp1}
\psi(r,\theta) &=\sum^\infty_{n=0}\frac{1}{n!}\left(\frac{r_0}{2}\lambda R\right)^n (\partial^{(n)}_r \psi)_0 = \psi_0(\theta) + \left(\frac{r_0}{2}R \psi_1(\theta) \right) \lambda + \mathcal{O}(\lambda^2),\\
\label{Iexp1}
I(r,\theta)&=\sum^\infty_{n=0}\frac{1}{n!}\left(\frac{r_0}{2}\lambda R\right)^n (\partial^{(n)}_r I)_0 = I_0(\theta) + \left(\frac{r_0}{2}R  I_1(\theta) \right) \lambda + \mathcal{O}(\lambda^2),\\
\label{Omegaexp1}
\Omega(r,\theta)&=\sum^\infty_{n=0}\frac{1}{n!}\left(\frac{r_0}{2}\lambda R\right)^n (\partial^{(n)}_r \Omega)_0 = \Omega_0(\theta) + \left(\frac{r_0}{2}R \Omega_1(\theta) \right) \lambda + \mathcal{O}(\lambda^2).
		\end{align}
\end{subequations}
We adopt the notation that $\psi_n=\psi_n(\theta) := (\partial^{(n)}_r \psi)_0$ is the $n$-th radial derivative of $\psi$ evaluated on the horizon of extreme Kerr spacetime. Similarly for $I$ and $\Omega$.
We assume in the following that $\psi_0'$ is different from zero. This is motivated by the fact that, for $\psi_0'=0$, the field $F^{(-1)}$ in Eq.~\eqref{Fm1} would vanish in the NHEK limit and the leading order field is electrically dominated, as discussed in Appendix \ref{App:B}.

The current vector $j$ has the following expansion for small $\lambda$ 
\begin{equation}
 j =j^\mu \partial_\mu= \sum_{n=-1}^{\infty}\lambda^{n}j^{(n)},
\end{equation}
with the leading-order term $j^{(-1)}$ given by 
\begin{equation}
\label{jm1}
j^{(-1)} = \frac{4}{r_0^3}\frac{1}{ \Gamma^2\Lambda}\left[\partial_{\theta}\left( \frac{\Lambda}{r_0}\left(r_0\Omega_0-1\right)\psi'_0\right)\left(\frac{\partial_T}{R^2}-\frac{\partial_\Phi}{R} \right) -I'_0 ~\partial_{R}\right].
\end{equation}
To leading order, the FF condition reads
\begin{equation}
    \left(F\cdot j\right)^{(-2)}=F^{(-1)} \cdot j^{(-1)} = \frac{2}{r_0^2 R^2}\frac{1}{\Lambda^2\Gamma^2}\partial_\theta\left[I^2_0-\frac{\Lambda^2}{r_0^2}(1-r_0\Omega_0)^2 (\psi'_0)^2\right]d\theta=0,
\end{equation}
where the dot means the contraction of the 2-form with the vector field.
To leading order in the $\lambda\rightarrow 0$ limit, the FF condition follows from imposing the Znajek condition \eqref{ZCExt}. 
This simplifies the expressions for $F^{(-1)}$ and $j^{(-1)}$, which now read 
\begin{equation}\label{FNHEK}
F^{(-1)}=\frac{r_0 I_0}{\Lambda} ~d\bigg(T-\frac{1}{R}\bigg) \wedge d\theta, \quad j^{(-1)} = \frac{4}{r_0^3}\frac{I'_0}{ \Gamma^2\Lambda}\left( \frac{\partial_T}{R^2}- \partial_{R} -\frac{\partial_\Phi}{R} \right).
\end{equation}
Notice that the Znajek condition allows to express the free function $I_0$ in terms of the two arbitrary functions $\psi_0$ and $\Omega_0$.
One can check that the field \eqref{FNHEK} is null $(F^{(-1)})^2=0$~\footnote{Since  $(F^{(-1)})^2=\left(F^2\right)^{(-2)}=\frac{8}{r^2_0 R^2}\frac{1}{\Lambda^2\Gamma^2}\left(I^2_0-\frac{\Lambda^2}{r_0^2}(1-r_0\Omega_0)^2 (\psi'_0)^2\right)$, one notices that the field being null, $(F^{(-1)})^2=0$, follows from the Znajek condition \eqref{ZCExt}.}, it obeys the Bianchi identity and it is regular on the future event horizon of the NHEK background. 
The scaling behavior of \eqref{FNHEK} is as expected, since under a rescaling $T \to T/c$ and $R \to cR$ the field strength (and the current vector) scales as $F^{(-1)}\to F^{(-1)}/c$, \emph{i.e.}, it is self-similar.
The solution \eqref{FNHEK} was found previously in \cite{Lupsasca:2014hua}. 
Furthermore, in \cite{Gralla:2016jfc}, it was shown that
any stationary, axisymmetric field, which is regular in extreme Kerr spacetime, has a limiting field that must be stationary, axisymmetric, null and self-similar. Here, we have concretely shown that the general stationary, axisymmetric, degenerate 2-form field \eqref{FKerr}, which is regular in extreme Kerr spacetime, converges to Eq.~\eqref{FNHEK}.
For this reason, we refer to this solution as the {\it NHEK attractor solution}~\cite{CGHOOletter}. 

It is interesting to remark that, once the NHEK limit is performed, the poloidal components of the magnetic field are sub-leading in $\lambda$ (namely, $F^{(-1)}\cdot \partial_\Phi=\mathcal{O}(\lambda)$); however, when higher orders in $\lambda$ are taken into account the poloidal magnetic field can appear.


\section{Post-NHEK corrections} \label{Sec:Climbing}

In the previous section, we considered the leading order FFE solution that one obtains by implementing the NHEK limiting procedure starting from a solution of FFE in the extreme Kerr background. This revealed that one ends up always with the attractor solution \eqref{FNHEK} in the NHEK limit, assuming a stationary, axisymmetric and regular field strength. In this section we move away from the NHEK attractor, in the sense that we want to perturbatively reconstruct a FF field $F$ in the extreme Kerr background by computing post-NHEK corrections to the FF field in the NHEK geometry.

We start again by considering the electromagnetic field \eqref{FKerr} in extreme Kerr spacetime with metric \eqref{KerrBL}. After moving to scaling coordinates \eqref{scaling} and expanding the field variables $\psi$, $\Omega$, $I$ around the event horizon as in Eq.~\eqref{expansions}, one obtains the formal expansions
\begin{subequations} \label{gFexpansions}
\begin{align}
g &= \sum_{n=0}^{\infty} \lambda^n g^{(n)}, \label{gexpansion}\\
F &= \sum_{n=-1}^{\infty} \lambda^n F^{(n)} \label{Fexpansion}
\end{align}
\end{subequations}
where $g^{(0)}_{\mu\nu}$ is the NHEK metric \eqref{NHEK} and $F^{(-1)}_{\mu\nu}$ is given by the attractor solution \eqref{FNHEK}.
The field strength $F$ obeys the FFE equations \eqref{Maxwell}-\eqref{FFE0}, which can be expanded as follows
\begin{subequations} \label{constraints}
\begin{align}
j &= \sum_{n=-1}^{\infty} \lambda^{n} j^{(n)}, \label{jexpansion}\\
dF &= \sum_{n=-1}^{\infty} \lambda^{n} (dF)^{(n)} =0, \label{Bianchiexpansion}\\
F\cdot j &= \sum_{n=-2}^{\infty} \lambda^{n} (F \cdot j)^{(n)} =0. \label{FFexpansion}
\end{align}
\end{subequations}
Here again the dot in $F \cdot j$ means the contraction $(F\cdot j)_\mu= F_{\mu\nu}j^{\nu}$.
Moreover, the Lorentz invariant $F^2=F^{\mu\nu}F_{\mu\nu}$ has the following expansion
\begin{equation} \label{F2expansion}
F^2 = \sum_{n=-2}^{\infty} \lambda^{n} (F^2)^{(n)}.
\end{equation}

A crucial point in the following will be that, even if the leading order FF field is null, $(F^2)^{(-2)}=0$, one can potentially obtain a magnetically-dominated FF field by going to higher orders in $\lambda$, which is highly relevant for astrophysical applications. Indeed, as seen in Sec.~\ref{Sec:Attractor}, the leading order FF field is null due to the Znajek condition \eqref{ZCExt}. However, a magnetically-dominated FF field around the extreme Kerr black hole is still possible by including the higher order corrections in $\lambda$. One of our main results below is that the first order for which $F^2$ can be non-zero is the second order, namely $(F^2)^{(0)}$; see Eq.~\eqref{F2exp} in Appendix~\ref{App:A}.

Let us summarize our procedure for finding the FF field $F$. 
At the leading order, namely the NHEK order ($n=-1$), we saw in Sec.~\ref{Sec:Attractor} that  $F^{(-1)}$ is given by the attractor solution \eqref{FNHEK}, which contains two arbitrary functions $\psi_0(\theta)$ and $\Omega_0(\theta)$~\footnote{This is a consequence of the Znajek condition \eqref{ZCExt} which allows one to express $I_0$ in terms of $\psi_0$ and $\Omega_0$.}; we assumed $\psi_0'$ is non-zero (see Appendix \ref{App:B} for the case with $\psi_0'=0$). $F^{(-1)}$ is regular at the event horizon, it obeys the Bianchi identity $(dF)^{(-1)}=0$ and it is null $(F^{(-1)})^2=0$, and the associated current $j^{(-1)}$ in Eq.~\eqref{FNHEK} is such that $F^{(-1)} \cdot j^{(-1)} =0$. 
To compute the higher order corrections $F^{(n)}$ in $\lambda$, \emph{i.e.} the post-NHEK corrections, one computes the general expressions of the field strength \eqref{Fexpansion} and the current \eqref{jexpansion} using the Kerr metric \eqref{KerrBL}, the Kerr field strength~\eqref{FKerr}, and the scaling coordinates~\eqref{scaling}. To this end, it is useful to expand also the metric components $g^{(i)}_{\mu\nu}$ as in Eq.~\eqref{gFexpansions}, as well as the Christoffel symbols $(\Gamma^{\mu}_{\nu \rho})^{(j)}$, to keep track of the orders in $\lambda$. The explicit expansions, up to the first orders in $\lambda$, of the field strength $F$, the current vector $j$, the FF constraint $F \cdot j$ and the invariant $F^2$ are relegated to Appendix~\ref{App:A}.

\subsection{1st post-NHEK order}\label{SubSec:1PNNHEK}

The 1st post-NHEK order of the field strength corresponds to $n=0$ and reads 
\begin{align} \label{F1PN}
F^{(0)} &= \frac{r_0^2}{2}\frac{I_0}{\Lambda}\frac{\psi_1}{\psi'_0}~dT\wedge dR + \frac{r_0^2}{2}\left(\frac{I_0}{\Lambda}\frac{\psi'_1}{\psi'_0} + \psi'_0 \Omega_1\right) RdT \wedge d\theta \nonumber\\
& +\frac{r_0}{2}\left(\frac{2}{\Gamma}\frac{I_0}{\Lambda}+r_0\frac{I_1}{\Lambda} \right)\frac{dR}{R}\wedge d\theta + \psi'_0 d\theta \wedge d\Phi.
\end{align}
As expected, the field is given in terms of the unknown field variables $\psi_1$, $\Omega_1$, and $I_1$ to be determined in terms of $\psi_0$ and $\Omega_0$ (recall that $I_0 = I_0(\psi_0, \Omega_0)$ from Eq.~\eqref{ZCExt}).
Notice that $F^{(0)}$ is scale-invariant under the rescalings $R \to cR$ and $T \to T/c$.
The expression for the current at this order is given by
\begin{align}\label{j1PN}
j^{(0)}  = j_{(0)}^{T}\frac{\partial_T}{R} +j_{(0)}^{R} R\partial_R+j_{(0)}^{\theta}\partial_\theta+j_{(0)}^{\Phi}\partial_\Phi ,
\end{align}
with
\begin{subequations} \label{j1PNcomponents}
\begin{align}
j_{(0)}^T &=\frac{2}{r_0^4}\frac{1}{\Gamma^2}\frac{I_0}{\Lambda}\left[\frac{\partial_\theta\left[(2+r_0^2\Omega_1)\Lambda \psi'_0 \right]}{I_0} +2r_0\Gamma\Lambda^2\left(\frac{1}{\Gamma^2\Lambda^2}\frac{\Gamma'}{\Gamma}-\frac{I'_0}{I_0} \right) + r_0^2\frac{\psi'_1}{\psi'_0}\left(\frac{I'_0}{I_0}-\frac{\psi''_0}{\psi'_0} + \frac{\psi''_1}{\psi'_1}\right)\right],\\
j_{(0)}^R &=\frac{2}{r_0^3}\frac{1}{\Gamma^2}\left(\frac{2}{\Gamma}\frac{I'_0}{\Lambda} - r_0\frac{I'_1}{\Lambda} \right),\\
j_{(0)}^\theta &=\frac{2}{r_0^2}\frac{1}{\Gamma^2}\frac{I_1}{\Lambda},\\
j_{(0)}^\Phi &=-j^{T} +\frac{2}{r_0^4}\frac{1}{\Gamma^2}\frac{I_0}{\Lambda}\left[ \frac{2}{\Lambda} \frac{\psi'_0}{I_0}\left(\frac{\psi''_0}{\psi'_0}-\frac{\Lambda'}{\Lambda}\right) + r_0 \Gamma^2\Lambda^2\left( \frac{\Lambda'}{\Lambda}+\frac{\Gamma'}{\Gamma}+\frac{1}{2}\frac{I'_0}{I_0}\right)- r_0^2\frac{\psi_1}{\psi'_0}\right].
\end{align}
\end{subequations}
The current $j^{(0)}$ is scale-invariant as well.

%
%
The Bianchi identity $(dF)^{(0)} =0$ implies the integrability condition $\psi_1\Omega'_0=\psi'_0\Omega_1$, whose solution is 
\begin{equation} \label{Omega1}
\Omega_1 = \frac{\Omega'_0}{\psi'_0}\psi_1.
\end{equation}
The FF condition $(F \cdot j)^{(-1)} =0$ implies that $\psi_1$ and $I_1$ must be given by
\begin{subequations}
\begin{align}
\psi_1 &= \frac{\psi'_0}{I'_0}I_1,\label{psi1} \\ 
I_1 &= \frac{\Lambda}{r_0}\left\{\partial_{\theta}\left[\left( r_0\Omega_0-1\right)\psi_1 \right]-\frac{\Lambda^2 \Gamma}{r_0} \left(r_0\Omega_0 - \frac{2}{\Gamma^2 \Lambda^2} \right)\psi_0'\right\}.\label{I1}
\end{align}
\end{subequations}
We emphasize that Eq.~\eqref{I1} is nothing but the supplemented Znajek condition for the regularity of the field at extremality \eqref{ZCExt2}, after substituting the field variables expansions \eqref{expansions} and using the result from the Bianchi identity in Eq.~\eqref{Omega1}. Remarkably, as we already noticed at the NHEK order, the regularity conditions \eqref{ZCExt} and \eqref{ZCExt2} of the field at the horizon and at extremality, automatically enforce the FF condition and the null condition $(F^2)^{(-1)}=0$.

From Eqs.~\eqref{psi1} and \eqref{I1}, we obtain a first-order linear differential equation for $\psi_1$ 
\begin{equation}
\psi'_1 -\left(\frac{\Lambda'}{\Lambda} + \frac{\psi''_0}{\psi'_0} \right)\psi_1 -\frac{\Lambda^2\Gamma}{r_0}\frac{\psi'_0}{r_0\Omega_0-1}\bigg(r_0\Omega_0-\frac{2}{\Lambda^2\Gamma^2}\bigg) = 0,
\end{equation}
from which one obtains $\psi_1= \psi_1(\psi_0, \Omega_0)$.
This allows one to write $\psi_1$, $\Omega_1$, and $I_1$ in terms of the functions $\psi_0$ and $\Omega_0$ as
\begin{subequations}\label{firstPNfunctions}
\begin{align}
\psi_1 &= \frac{\mathcal{G}\Lambda}{r_0}\psi'_0,\\
\Omega_1 &= \frac{\mathcal{G}\Lambda}{r_0}\Omega'_0,\\
I_1 &= \frac{\mathcal{G}\Lambda}{r_0} I'_0 = \frac{\mathcal{G}\Lambda}{r_0}\partial_{\theta}\left[\frac{\Lambda}{r_0}\left(r_0\Omega_0-1\right)\psi'_0 \right],
\end{align}
\end{subequations}
where the function $\mathcal{G}$ is defined by
\begin{equation}
\label{calG}
\mathcal{G'}=\frac{\Lambda\Gamma}{r_0\Omega_0-1}\bigg(r_0\Omega_0-\frac{2}{\Lambda^2\Gamma^2}\bigg).
\end{equation}
We notice that the function $\mathcal{G}$  simplifies to a constant if $r_0\Omega_0=2/(\Lambda^2\Gamma^2)$. As we shall see below in Sec.~\ref{SubSec:MD}, this condition is obeyed by the only known exact solution to FFE in Kerr spacetime: the Menon-Dermen solution \cite{MD2005,Menon_2011}. 

To summarize, the field strength \eqref{F1PN} simplifies to 
\begin{align}
F^{(0)} &= \frac{r_0}{2}\mathcal{G}I_0~dT\wedge dR + \frac{r_0}{2}\mathcal{G}\left[I_0\left(\frac{\mathcal{G'}}{\mathcal{G}} + \frac{\Lambda'}{\Lambda} + \frac{\psi''_0}{\psi'_0}\right) + \Lambda \psi'_0 \Omega'_0\right] RdT \wedge d\theta \nonumber\\
& + \frac{r_0}{2} \left(\frac{2I_0}{\Gamma \Lambda} + \mathcal{G}I'_0\right)~\frac{dR}{R}\wedge d\theta + \psi'_0 d\theta \wedge d\Phi,
\end{align}
with current given by Eqs.~\eqref{j1PN}-\eqref{j1PNcomponents} and with $\psi_1$, $\Omega_1$ and $I_1$ as in Eq.~\eqref{firstPNfunctions}.

\subsection{2nd post-NHEK order} \label{SubSec:2PNNHEK}

The next order in the post-NHEK expansion of the field strength (corresponding to $n=1$) gives 
\begin{align} \label{F2PN}
F^{(1)} &= \frac{r_0^3}{4}\left(\frac{I_0}{\Lambda}\frac{\psi_2}{\psi'_0} + \psi_1\Omega_1 \right)RdT\wedge dR +\frac{r_0^3}{8}\left(\frac{I_0}{\Lambda}\frac{\psi'_2}{\psi'_0} + \psi'_0\Omega_2 +2\psi'_1\Omega_1 \right)R^2dT\wedge d\theta \nonumber\\
& +\frac{r_0}{8}\left(\frac{4}{\Gamma}\frac{I_0+r_0 I_1}{\Lambda} + r_0^2\frac{I_2}{\Lambda} \right)dR\wedge d\theta + \frac{r_0}{2}\psi_1 dR\wedge d\Phi +\frac{r_0}{2}\psi'_1 Rd\theta \wedge d\Phi,
\end{align}
with $\psi_1$, $\Omega_1$ and $I_1$ given in Eq.~\eqref{firstPNfunctions} and with $\psi_2$, $\Omega_2$ and $I_2$ to be determined in terms of $\psi_0$ and $\Omega_0$ by solving the equations of motion at this order. 
Under the scalings $T \to T/c$ and $R \to cR$, we observe that $F^{(1)} \to c F^{(1)}$, as expected. The current vector at the 2nd post-NHEK order is lengthy and it is written in Appendix \ref{Expr2ndNHEK}.

From the Bianchi identity, $(dF)^{(1)}=0$, we get the condition
\begin{equation}
\label{Omega2}
\Omega_2 = \frac{\Omega'_0}{\psi'_0}\psi_2 +\frac{\Omega'_1}{\psi'_0}\psi_1+\frac{\psi'_1}{\psi'_0}\Omega_1 = \left[\frac{\psi_2}{\psi'_0} + \frac{\mathcal{G}^2\Lambda^2}{r_0^2}\left(\frac{\Omega''_0}{\Omega'_0} - \frac{\psi''_0}{\psi'_0}\right) \right]\Omega'_0.
\end{equation}
The FF condition, $(F \cdot j)^{(0)} =0$, instead, implies the following functional form for $I_2$ 
\begin{equation}
\label{I2}
I_2 = \frac{I'_0}{\psi'_0}\psi_2 +\frac{I'_1}{\psi'_0}\psi_1+\frac{\psi'_1}{\psi'_0}I_1 = \left[\frac{\psi_2}{\psi'_0} + \frac{\mathcal{G}^2\Lambda^2}{r_0^2}\left(\frac{I''_0}{I'_0} - \frac{\psi''_0}{\psi'_0}\right) \right]I'_0,
\end{equation}
and a second-order linear differential equation for $\psi_2$ 
\begin{equation}
\label{psi2_eq}
    \psi''_2 + a(\theta)\psi'_2 + b(\theta)\psi_2 + c(\theta) =0, 
\end{equation}
with coefficients given by
\begin{subequations} \label{psi2_coeffs}
\begin{align}
    a(\theta) &=  2\frac{I'_0}{I_0}- \frac{\Lambda'}{\Lambda} -2 \frac{\psi''_0}{\psi'_0},\\
    b(\theta) &= 2 - \frac{\Lambda''}{\Lambda}- \frac{\psi^{(3)}_0}{\psi'_0} + 2 \left(\frac{\psi''_0}{\psi'_0}\right)^2 - \left(2\frac{I'_0}{I_0} -\frac{\Lambda'}{\Lambda} \right)\left(\frac{\Lambda'}{\Lambda} +\frac{\psi''_0}{\psi'_0}\right),\\
    c(\theta) &= -\frac{\Lambda^2 \psi'_0}{r_0^4}\left(A(\theta) + \mathcal{G}(\theta) B(\theta) +\mathcal{G}^2(\theta) C(\theta)\right),
\end{align}
\end{subequations}
where the functions $A(\theta)$, $B(\theta)$ and $C(\theta)$ are explicitly written in Appendix \ref{Expr2ndNHEK}.

Finally, the Lorentz invariant at this order is given by
\begin{equation}\label{F2}
(F^2)^{(0)} = \frac{2 }{r_0^4}\left(\frac{\psi'_0}{\Gamma\Lambda} \right)^2\left(D(\theta) + \mathcal{G}(\theta)E(\theta)\right),
\end{equation}
where the expressions for $D(\theta)$ and $E(\theta)$ can be found in Appendix \ref{Expr2ndNHEK}.
We will not go beyond the 2nd post-NHEK order in this paper. In the following sections we will instead take advantage of the analytical expressions we have obtained so far to derive, in a new way, the well-known Menon-Dermer class of solutions in Sec.~\ref{SubSec:MD} and to construct novel solutions in Sec.~\ref{SubSec:Ansatz 1}.



\section{Menon-Dermer class from the NHEK order} 
\label{SubSec:MD}

To date the only known exact stationary and axisymmetric solution to FFE in Kerr background is the Menon-Dermer class of solutions \cite{MD2005,Menon_2011}. These solutions are represented by a set of field variables $(\psi_{\text{MD}}(\theta), I_{\text{MD}}(\theta), \Omega_{\text{MD}}(\theta))$ that do not depend on the radial coordinate; in particular $r_0\Omega_{\text{MD}}=2/\sin^2\theta$ is fixed, whereas $I_{\text{MD}}(\theta)$ is specified by the Znajek condition once an arbitrary stream function $\psi_{\text{MD}}(\theta)$ has been chosen. The current associated to this solution flows along the principal null geodesics of Kerr and, as a consequence, the magnetosphere is everywhere null, namely $F^2_{\text{MD}}=0$.
Generalizations of this class to time-dependent and non-stationary cases were constructed in \cite{BGJ}, exploiting the principal null congruence in Kerr as an ansatz to solve the FF constraint.

In this section, we show that the Menon-Dermer class follows from the condition that all post-NHEK orders in the $\lambda$-expansion \eqref{expansions} are set to zero; as a matter of fact, demanding that $(\psi_n, \Omega_n, I_n)$ should vanish $\forall n\geq1$ gives no dependence on the radial coordinate for the extreme Kerr field variables $(\psi, I, \Omega)$.
Indeed, the field angular velocity of the Menon-Dermer class can be derived explicitly by demanding that the 1st post-NHEK order \eqref{firstPNfunctions} vanishes. This is achieved by requiring that the function $\mathcal{G}$ in Eq.~\eqref{calG} should be zero, which precisely selects
\begin{equation}
\label{Omega_MD}
r_0\Omega_0=\frac{2}{\Lambda^2 \Gamma^2}, 
\end{equation}
corresponding to $\Omega_{\text{MD}}$. Under this condition, the polar currents \eqref{ZCExt} reads 
\begin{equation}
\label{I0_MD}
I_0=\frac{2}{r_0}\frac{\psi'_0}{\Lambda \Gamma},
\end{equation}
where $\psi_0$ remains an arbitrary function. 

Assuming that the 2nd and higher post-NHEK orders vanish, one can directly write the extreme Kerr field \eqref{FKerr} in terms of the NHEK field variables $\psi_{\text{MD}}\equiv\psi_0$, $\Omega_{\text{MD}}\equiv\Omega_0$ and $I_{\text{MD}}\equiv I_0$. Using Eq.~\eqref{I0_MD}, it is possible to rearrange the field strength \eqref{FKerr} as follows 
\begin{equation}
\label{F_MD}
F_{\text{MD}}=-\frac{I_0}{\Lambda \Gamma}d\theta\wedge \left[dt+\frac{r^2+(r_0/2)^2}{\Delta}dr\right]+\frac{r_0}{2}\Lambda \Gamma I_0 d\theta\wedge\left(d\phi+\frac{r_0}{2\Delta}dr\right),
\end{equation}
which is regular on the future event-horizon\footnote{Indeed, the quantities in the square brackets are nothing but the ingoing Kerr coordinates $1$-forms $$dv=dt+[(r^2+a^2)/\Delta] dr,\quad d\bar{\phi}=d\phi+(a/\Delta) dr,$$
when extremality is reached. This allows us to write \eqref{F_MD} in the typical form of an ingoing flux in Extreme Kerr, $F_{MD}=-I_0\left(\Lambda\Gamma\right)^{-1} d\theta\wedge(dv-\Omega_0^{-1}d\bar{\phi})$.} and leads to $F^2_{\text{MD}}=0$.
The NHEK field strength, together with its associated current, are still given by Eq.~\eqref{FNHEK}.
We notice that the extreme Kerr field \eqref{F_MD}, as well as its NHEK limit and its associated current, appear to be singular on the rotational axis; this kind of singularity, unlike the divergence of $\Omega_0$ in \eqref{Omega_MD}, is not an intrinsic feature of the MD class: the function which allows to distinguish an MD solution from another is $\psi_0$, and for every $\psi_0=h(\theta) \Lambda^{k+1}$, with $k\geq2$ and $h(\theta)$ some arbitrary function regular on the axis, the field turns out to be regular \cite{Menon_2011,MDbook}.

The vector current associated to \eqref{F_MD} can be written as 
\begin{equation}
\label{JCal_MD}
j^\mu(r,\theta)=\frac{\mathcal{J}(\theta)}{2\Sigma}n^\mu, \quad \mathcal{J}(\theta):=\frac{2I_0'}{\Lambda \Gamma}=\frac{4}{r_0}\frac{\psi'_0}{\Lambda^2\Gamma^2}\left(\frac{\psi_0''}{\psi'_0} - \frac{\Lambda'}{\Lambda} -\frac{\Gamma'}{\Gamma}\right),
\end{equation}
with the vector
\begin{equation}
n=\left(\frac{(r^2+a^2)}{\Delta}\partial_t-\partial_r+\frac{a}{\Delta}\partial_\phi\right)\bigg|_{a=r_0/2}
\end{equation}
which identifies the principal ingoing null geodesic in extreme Kerr.\footnote{By choosing the opposite sign in Eq.~\eqref{ZCExt}, one obtains a field strength which is regular on the past event-horizon and whose vector current lies along the principal outgoing null geodesic.} 
This is a crucial signature of the MD solutions since, as proven in \cite{BGJ}, this class contains all the stationary, axisymmetric FF fields with vector currents along the infalling principal null direction.

\section{Novel perturbative solutions}
\label{SubSec:Ansatz 1}

Any regular, stationary and axisymmetric FF field in the background of extreme Kerr reduces to the attractor solution \eqref{FNHEK} with a null field strength. As we saw in Sec.~\ref{SubSec:MD}, this includes the Menon-Dermer class of solutions for which the field strength is null everywhere. The question that we address in the following is whether it is possible to use the tools we developed in Sec.~\ref{Sec:Climbing} to move away from the NHEK attractor to a magnetically-dominated solution.

To construct solutions that are not in the Menon-Dermer class, one has to take into account post-NHEK orders. To this aim, we have computed the 1st and 2nd post-NHEK orders in Sec.~\ref{Sec:Climbing}. To summarize the results of Sec.~\ref{Sec:Climbing}, while the field variables of the 1st post-NHEK order can be easily recast in a simple form (see Eq.~\eqref{firstPNfunctions}) in terms of 
$(\psi_0, I_0, \Omega_0)$, the 2nd post-NHEK order is more involved. To derive the field strength at the 2nd post-NHEK order, one needs to solve the second-order linear differential equation \eqref{psi2_eq} with coefficients \eqref{psi2_coeffs} given by the arbitrary NHEK functions $\psi_0$ and $\Omega_0$ and their derivatives. 

In order to analytically solve Eq.~\eqref{psi2_eq}, we consider the following ansatz for $\psi_0$ 
\begin{equation}
\label{Ansatz1}
\frac{\psi_0''}{\psi_0'}=-\frac{3}{2}\left(\frac{\Lambda'}{\Lambda}\right)+\frac{r_0\Omega'_0}{1-r_0\Omega_0},
\end{equation}
or equivalently, upon integration (assuming $\psi_0(0) = 0$),
\begin{equation} \label{Ansatz1bis}
    \psi_0(\theta) = k_0 \int (1-r_0\Omega_0)^{-1} \Lambda^{-3/2}d\theta.
\end{equation}
With the ansatz \eqref{Ansatz1}, the differential equation \eqref{psi2_eq} becomes
\begin{equation}
\label{58_Polyanin}
\psi_2''+a(\theta)\psi_2'+\left(\frac{a'(\theta)}{2}+\frac{a^2(\theta)}{4}+2\right)\psi_2 + c(\theta)=0,
\end{equation} 
with $\Omega_0$ arbitrary and $a(\theta)$ and $c(\theta)$ given in Eq.~\eqref{psi2_coeffs}.
The most general solution of Eq.~\eqref{58_Polyanin} is given by \cite{polyanin2003handbook}
\begin{equation}\label{psi2full}
\psi_2(\theta)=\frac{1}{\left(1-r_0\Omega_0\right)\Lambda^{1/2}}\left[\psi^{h}_2(\theta)+\psi_2^{nh}(\theta)\right],
\end{equation}
where the homogeneous and non-homogeneous solutions are, respectively, 
\begin{subequations}
\begin{align}\label{psi2hom}
\psi^{h}_2(\theta)&=c_1\cos\left(\sqrt{2}\theta\right)+c_2\sin\left(\sqrt{2}\theta\right), \quad c_1, c_2 \in \mathbb{R},\\\label{psi2nonhom}
\psi_2^{nh}(\theta)&= +\cos\left(\sqrt{2}\theta\right)\int c(\theta)~\left(1-r_0\Omega_0\right)\Lambda^{1/2}~\frac{\sin\left(\sqrt{2}\theta\right)}{\sqrt{2}}d\theta \nonumber\\
&\quad -\sin\left(\sqrt{2}\theta\right)\int c(\theta)~\left(1-r_0\Omega_0\right)\Lambda^{1/2}~\frac{\cos\left(\sqrt{2}\theta\right)}{\sqrt{2}}d\theta,
\end{align}
\end{subequations}
with $c(\theta)$ explicitly given in Eq.~\eqref{psi2_coeffs}.

Since the angular velocity $\Omega_0$ is arbitrary, one can either choose it equal to $\Omega_{\text{MD}}$ (see Eq.~\eqref{Omega_MD}) and, starting from that, construct radial corrections to the Menon-Dermer class, or one can choose a different function and construct novel perturbative solutions. 
As an educated guess for the arbitrary function $\Omega_0$, we introduce the following class of angular velocities parametrized by $\beta \neq 0$
\begin{equation}\label{OmegaAnsatz}
    r_0 \Omega_0 = 1 +\frac{\beta}{2}\left( 1- \frac{2}{\Gamma^2\Lambda^2}\right), \quad \beta \in \mathbb R_{\ne 0},
\end{equation}
from which it turns out that (see Eq.~\eqref{calG}) 
\begin{equation} \label{calGAnsatz}
    \mathcal{G}(\theta)= g - \left(1+\frac{2}{\beta}\right)\cos(\theta),
\end{equation}
where $g$ is an integration constant.
We notice that the particular choice of the field angular velocity \eqref{OmegaAnsatz} with $\beta=-2$ amounts to the angular velocity of the Menon-Dermer class of solutions (see Eq.~{\eqref{Omega_MD}}). The ansatz \eqref{Ansatz1} for $\psi_0$, then, selects a specific solution within this class. 

In the following, we are going to compute the NHEK, 1st and 2nd post-NHEK orders for arbitrary $\beta$ and $g$.
Given the angular velocity \eqref{OmegaAnsatz}, we compute $\psi_0$ from Eq.~\eqref{Ansatz1bis} and $I_0$ from the Znajeck condition \eqref{ZCExt}. Thus, the NHEK order is given by \footnote{Notice that since $I_0$ is singular at the rotation axis, thus the field strength $F$ will also be singular.}
\begin{subequations}
\begin{align}
\psi_0 &=\frac{k_0}{\beta}\int \Gamma\Lambda^{1/2}d\theta,\\ I_0 &=-\frac{k_0}{r_0}\frac{1}{\Lambda^{1/2}},\\ 
r_0\Omega_0 &=1 +\frac{\beta}{2}\left( 1- \frac{2}{\Gamma^2\Lambda^2}\right).
\end{align}
\end{subequations}
The 1st post-NHEK order is then given in Eq.~\eqref{firstPNfunctions} and it now reads
\begin{subequations}
    \begin{align}
        \psi_1 &= \frac{1}{\beta^2}\frac{k_0}{r_0}\Gamma\Lambda^{3/2}\left[\beta g + (\beta+2)\int \Gamma \Lambda~ d\theta \right],\\
        I_1 &=\frac{1}{2\beta}\frac{k_0}{r_0^2}\frac{\Lambda'}{\Lambda^{1/2}}\left[\beta g + (\beta+2)\int \Gamma \Lambda~ d\theta \right],\\
        \Omega_1 &= \frac{2}{r_0^2}\frac{1}{\Gamma^2 \Lambda}\left(\frac{\Lambda'}{\Lambda} +\frac{\Gamma'}{\Gamma} \right)\left[\beta g + (\beta+2)\int \Gamma \Lambda~ d\theta \right].
    \end{align}
\end{subequations}
The 2nd post-NHEK order, in terms of $\psi_2$ and the NHEK functions $(\psi_0, I_0, \Omega_0)$, is (see Eqs.~\eqref{Omega2} --\eqref{I2})
\begin{subequations} \label{I2Omega2}
\begin{align}
I_2 &=  \left[\frac{\psi_2}{\psi'_0} + \frac{\mathcal{G}^2\Lambda^2}{r_0^2}\left(\frac{I''_0}{I'_0} - \frac{\psi''_0}{\psi'_0}\right) \right]I'_0,\\
\Omega_2 &=  \left[\frac{\psi_2}{\psi'_0} + \frac{\mathcal{G}^2\Lambda^2}{r_0^2}\left(\frac{\Omega''_0}{\Omega'_0} - \frac{\psi''_0}{\psi'_0}\right) \right]\Omega'_0.
\end{align}
\end{subequations}
The stream function $\psi_2$ can be read in Eqs.~\eqref{psi2full}-\eqref{psi2hom}-\eqref{psi2nonhom}, while $\mathcal{G}$ is given in Eq.~\eqref{calGAnsatz}. It is interesting to mention that $\psi_2 = \mathcal{O}(\theta^2)$, so it is regular on the rotation axis. Moreover, for the special choice $(\beta=-2, g=0)$, which amounts to the Menon-Dermer field angular velocity, the non-homogeneous part $\psi_2^{nh}$ vanishes as well as the 1st post-NHEK order. We will further notice that, after fixing the two coefficients $c_1$ and $c_2$ as in Eq.~\eqref{c1c2}, also $\psi_2^{h}$ vanishes and so $(I_2, \Omega_2)$ do.

\begin{figure}[h!]
    \centering
    \includegraphics[scale=1.2]{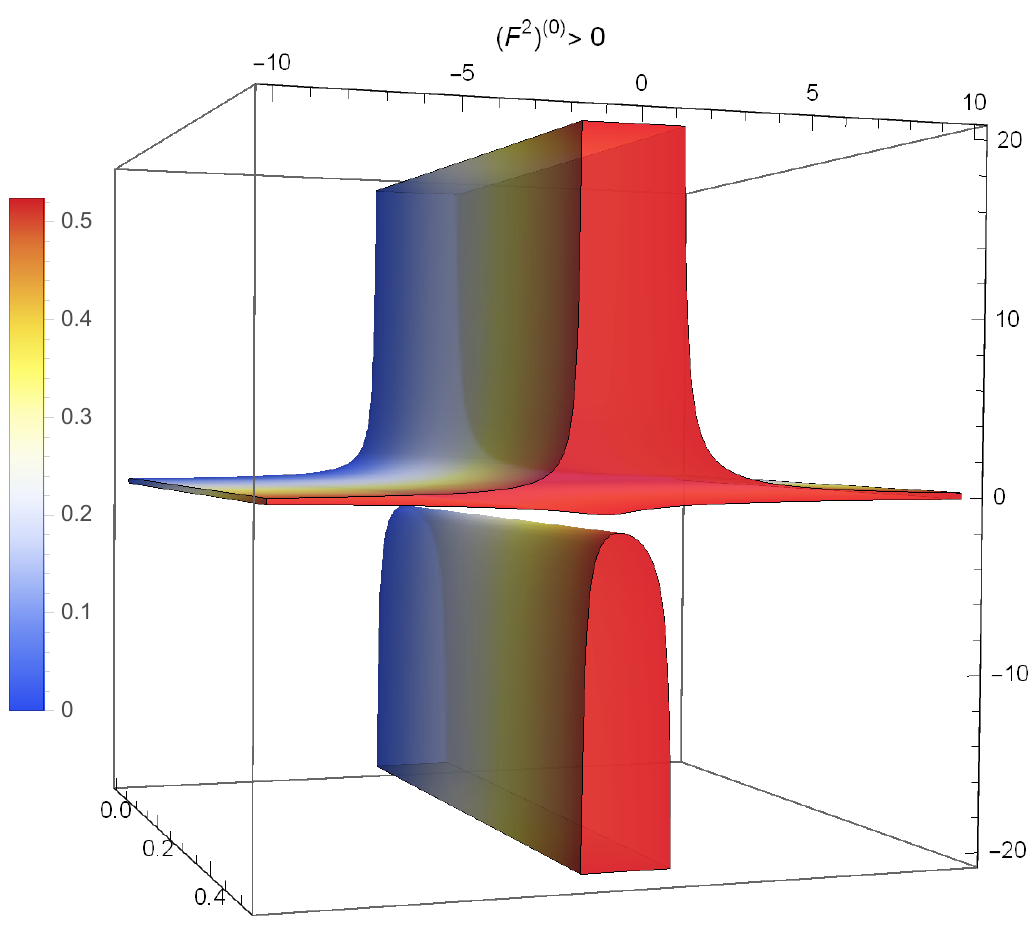}
    \caption{We consider the Taylor expansion of $(F^2)^{(0)}$ in Eq.~\eqref{TaylorF2general} with $r_0 =1$, $k_0=2$, and coefficients $c_1$, $c_2$ as in Eq.~\eqref{c1c2}. We plot the positive region of $\left(F^2\right)^{(0)}$ for parameters in the ranges $g \in (-10,+10)$ and $\beta \in (-20,+20)$.}
    \label{fig:RegionPositiveF2}
\end{figure}

\begin{figure}[h!]
    \centering
    \includegraphics[scale=1.2]{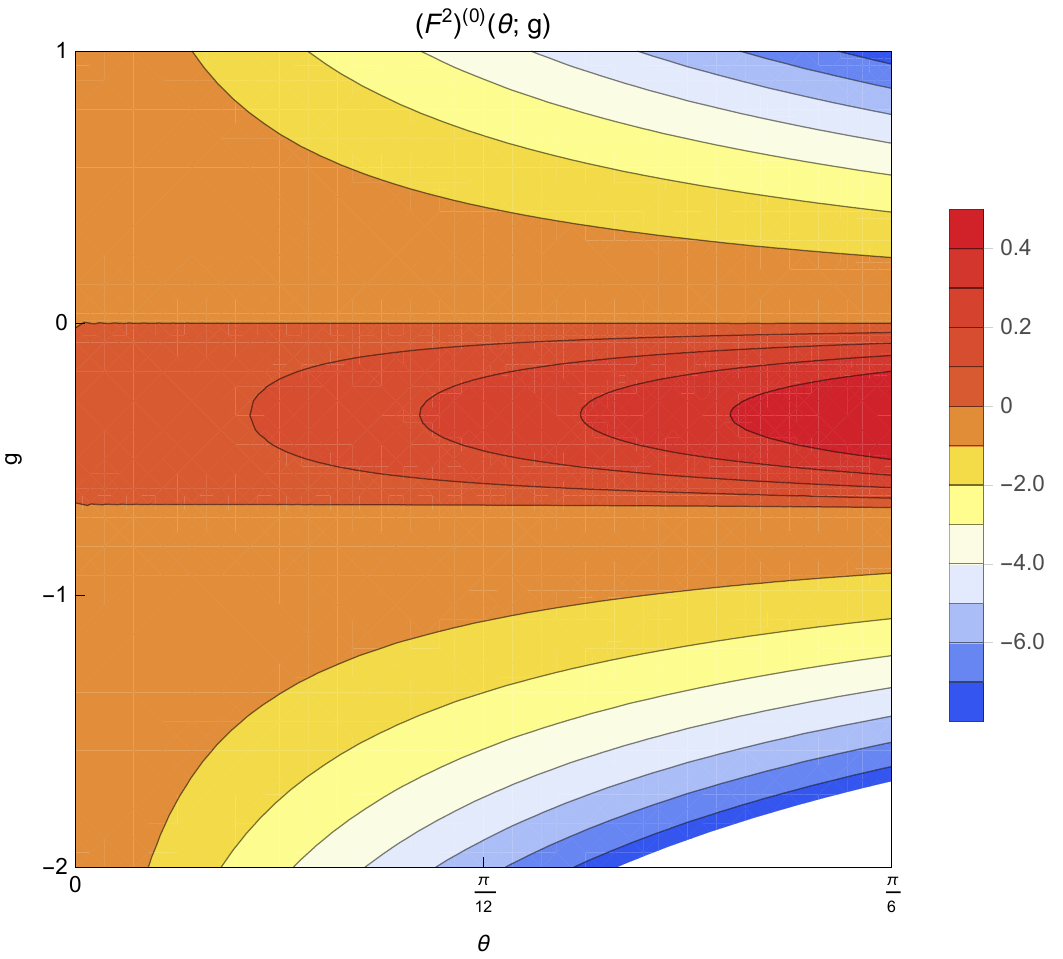}
    \caption{We consider the Taylor expansion of $(F^2)^{(0)}$ in Eq.~\eqref{TaylorF2general} with $r_0 =1$, $k_0=2$, $\beta=-2$, $c_1=-g(4-5g)$, and $c_2=0$. The middle strip, defined by $-0.67\lessapprox  g<0$, is the range of values for which $(F^2)^{(0)}$ is positive, \emph{i.e.}, the field strength is a magnetically-dominated solution to FFE.}
    \label{fig:F2contour}
\end{figure}

An important question to investigate is the sign of the Lorentz invariant $(F^2)^{(0)}$ in Eq.~\eqref{F2}. The Menon-Dermer class of solutions is null, \emph{i.e.}, $F^2=0$. Radial contributions to this class, however, can change the character of the FF solution from null to either electrically or magnetically-dominated corresponding, respectively, to $F^2<0$ and $F^2>0$. To analytically study the function $(F^2)^{(0)}$ everywhere is not an easy task, because the non-homogeneous part of $\psi_2$, given by Eq.~\eqref{psi2nonhom}, involves a difficult integral. 
However, the expression for $(F^2)^{(0)}$ admits the following Taylor expansion around the rotation axis 
\begin{align}
    \left(F^2\right)^{(0)} &= -2\sqrt{2}\frac{k_0}{r_0^2}\frac{c_2}{\theta^2} \nonumber\\
    &\quad + 2\frac{k_0}{r_0^2}\left[2c_1 -5\frac{k_0}{r_0^2} g^2 + 2 \frac{k_0}{r_0^2} \left(7+\frac{10}{\beta}\right)g - \frac{k_0}{r_0^2}\frac{(2+\beta)(6+11\beta)}{\beta^2} \right]\frac{1}{\theta}\\
    &\quad +\frac{4\sqrt{2}}{3}\frac{k_0}{r_0^2}c_2 + \mathcal{O}(\theta).\nonumber
\end{align}
Our goal is to find a regular expression for $\theta \to 0$; this amounts to set the arbitrary coefficients in the stream function $\psi_2$ (see Eq.\eqref{psi2hom}) to be
\begin{equation} \label{c1c2}
    c_1 = \frac{1}{2}\frac{k_0}{r_0^2}\left[ 5g^2 - 2\left(7+\frac{10}{\beta}\right)g+ \frac{(2+\beta)(6+11\beta)}{\beta^2}\right], \quad c_2=0.
\end{equation}
The condition $c_2=0$ sets to zero all coefficients of the even powers in the Taylor expansion. 
Then, the first contributions for small polar angles are given by
\begin{align} \label{TaylorF2general}
    \left(F^2\right)^{(0)} &= +\left[-8g^2 + \frac{32}{3}\frac{g}{\beta} + \frac{16}{3}\frac{(2+\beta)(2+3\beta)}{\beta^2}\right]\theta \nonumber\\
    &\quad - \left[\frac{34}{15}g^2 + \left(4+\frac{40}{9}\frac{1}{\beta} \right)g -\frac{8}{45}\frac{(2+\beta)(11+12\beta)}{\beta^2} \right]\theta^3\nonumber\\
    &\quad +\left[\frac{g^2}{21} -\frac{8}{105}\frac{43+25\beta}{\beta}g + \frac{8}{105}\frac{(2+\beta)(6+\beta)}{\beta^2} \right]\theta^5\nonumber\\
    &\quad +\left[\frac{349}{1400}g^2-\frac{3770+1767\beta}{5670}\frac{g}{\beta} - \frac{1}{56700}\frac{(2+\beta)(18968+13581\beta)}{\beta^2}\right]\theta^7\nonumber\\
    &\quad +\left[\frac{203537}{2494800}g^2+\frac{3398+2348\beta}{31185}\frac{g}{\beta}-\frac{(2+\beta)(187794+100163\beta)}{1247400\beta^2}\right]\theta^9 \nonumber\\
    &\quad+ \mathcal{O}(\theta^{11}).
\end{align}
Fig.~\ref{fig:RegionPositiveF2} shows the pairs $(\beta, g)$ for which $\left(F^2\right)^{(0)}(\theta)$ is positive in a neighborhood of the axis of rotation.
For the sake of concreteness, let us choose $\beta=-2$. As mentioned earlier, this choice amounts to consider the Menon-Dermer field angular velocity. The simplest choice $g=0$ implies that $(F^2)^{(0)}=0$.
For $g\neq 0$, \emph{i.e.}, when radial contributions to the Menon-Dermer solution are taken into account, the Taylor expansion of $(F^2)^{(0)}$ simplifies and its values are plotted in Fig.~\ref{fig:F2contour}. The middle strip, depicted in Fig.~\ref{fig:F2contour} and defined by $-0.67\lessapprox  g<0$, is the intersection of the three-dimensional region in Fig.~\ref{fig:RegionPositiveF2} with the plane defined by $\beta=-2$. Any choice of $-0.67\lessapprox  g<0$ guarantees that $(F^2)^{(0)}$ is positive and therefore the field strength is magnetically-dominated. 

To conclude, let us compute the energy and angular momentum extraction from Eq.~\eqref{EJextraction}. At the leading order in $\lambda$, the outflows of energy and angular momentum are
\begin{subequations}\label{EJextractionNHEK}
\begin{align}
    \frac{1}{2\pi}\frac{dE}{dt} &= \int \Omega_0\left(\frac{1}{r_0} - \Omega_0 \right)\left(\psi'_0 \right)^2\Lambda~ d\theta \nonumber\\
    &= \frac{k_0^2}{8\beta r_0^2}\left[(6+7\beta)\theta + \beta(8\cot \theta +\cos \theta \sin \theta)+\sin(2\theta)\right],\\
     \frac{1}{2\pi}\frac{dJ}{dt} &= \int \left(\frac{1}{r_0} - \Omega_0 \right)\left(\psi'_0 \right)^2\Lambda~ d\theta \nonumber\\
     &= \frac{k_0^2}{8\beta r_0}\left[6\theta + \sin(2\theta)\right],
\end{align}
\end{subequations}
where in the second equality we used the ansatz \eqref{Ansatz1} for $\psi_0$ and \eqref{OmegaAnsatz} for $\Omega_0$. While the angular momentum outflow is always finite and negative for $\beta <0$, the energy outflow diverges at the rotation axis. It is a consequence of the ansatzes \eqref{Ansatz1} and \eqref{OmegaAnsatz}, responsible also of the divergence of the field strength at the rotation axis. We leave for future investigation the searching of a regular magnetically-dominated field strength, with finite energy outflow. These regularity issues could be cured by taking into account the presence of the inner light-surface that separates the near-horizon and the near-axis regions.

\section{Summary }
\label{Sec:Conclusion}

In this paper, following the approach introduced in our letter \cite{CGHOOletter}, we have proposed a perturbative procedure to construct stationary, axisymmetric, FF magnetospheres around extreme Kerr black holes that are magnetically-dominated. Our approach, as well as the results in this paper, are analytical; however, as already discussed in the introduction, it would be interesting to numerically implement the perturbative algorithm to higher orders. 

Let us summarize and comment the main results of the paper.
We first reviewed in Sec.~\ref{Sec:Attractor} the NHEK attractor solution \eqref{FNHEK} defined in the NHEK geometry, that is the limit of any stationary, axisymmetric, FF field strength \eqref{FKerr} which is regular in extreme Kerr spacetime. In other words, regardless of the field defined in extreme Kerr spacetime, with the above features, one always ends up in the NHEK limit with the attractor solution \eqref{FNHEK}. This observation is the starting point of our perturbative scheme. The NHEK attractor solution has a precise tensor structure, dictated by the global conformal symmetry of the background geometry. However, it contains two arbitrary functions in its components.

In Sec.~\ref{Sec:Climbing}, we outlined the general procedure to construct post-NHEK orders around the NHEK attractor solution. Pictorially, the expansion in the parameter $\lambda$ amounts to move away from the NHEK attractor solution towards solutions of FFE in extreme Kerr spacetime. The motivation behind this programme is to show that the invariant $F^2$ gets corrections and eventually, at a certain post-NHEK order, the FF field is magnetically-dominated.

To achieve this result, we have explicitly computed the 1st and 2nd post-NHEK orders summarized, respectively, in Eq.~\eqref{firstPNfunctions}
and Eqs.~\eqref{Omega2}-\eqref{I2}-\eqref{psi2_eq}.
The first technical result is the derivation of the second-order linear differential equation for the 2nd post-NHEK stream function \eqref{psi2_eq}, with coefficients \eqref{psi2_coeffs} depending on the arbitrary functions present in the NHEK attractor solution. En passant, we recovered the well-known Menon-Dermer class of solutions in Sec.~\eqref{SubSec:MD} by demanding the vanishing of all post-NHEK orders.
The second technical result, presented in Sec.~\ref{SubSec:Ansatz 1}, is the analytic solution of the differential equation \eqref{psi2_eq}, by providing the ansatz \eqref{Ansatz1} that relates the arbitrary functions in the NHEK attractor solution.
The ansatz has been pivotal to compute the field strength up to the 2nd post-NHEK order and show that $F^2$, after being regularized at the rotation axis, can be positive. This result is obtained in Eq.~\eqref{TaylorF2general} and in Figs.~\ref{fig:RegionPositiveF2}-\ref{fig:F2contour}.
However, despite the analytic solution and the promising result that magnetically-dominated FF solutions can be constructed perturbatively with finite angular momentum outflow, the ansatz aforementioned has the drawback that the field strength as well as the energy outflow are not regular on the rotation axis. 

As a side result, in Appendix~\ref{App:B}, we have also found a new NHEK solution (see Eq.~\ref{newnhek}) that is scale-invariant and electrically-dominated. It is the most general NHEK attractor solution in the case where $\psi_0$ is constant.

There are several interesting directions to continue these investigations, as also mentioned in \cite{CGHOOletter}. One important aspect to investigate is the role of light-surfaces that appear close to the event horizon. The inner light-surface is of particular interest as it separates the near-horizon region and the region close to the axis. An analysis of this issue is highly relevant to understand possible singular behavior near the rotation axis.
It would also be very interesting to study numerically the differential equation \eqref{psi2_eq} with physical boundary conditions to make the field strength regular on the axis. This might provide regular solutions that could be very relevant for astrophysical applications. 
Another important direction is to generalize the methods of this paper to the near-NHEK limit. In \cite{CGHOOletter}, the first step has already been taken by finding the near-NHEK attractor solution. Following the current work, one should develop the perturbation theory away from the near-NHEK limit, in a similar fashion to the method of this paper for the NHEK limit. This could reveal whether one can also find solutions with positive $F^2$ in this case.

\paragraph{Acknowledgements}
We thank G.~Comp\`ere, V.~Karas, G.~Menon, and the anonymous referee for interesting comments and feedback.
T.~H.~acknowledges support from the Independent Research Fund Denmark grant number DFF-6108-00340 ``Towards a deeper understanding of black holes with non-relativistic holography". G.~G. and M.~O.~acknowledge support from the project ``Black holes, neutron stars and gravitational waves" financed by Fondo Ricerca di Base 2018 of the University of Perugia.
R.O. is funded by the European Structural and Investment Funds (ESIF) and the Czech Ministry of Education, Youth and Sports (MSMT), Project CoGraDS - CZ.02.1.01/0.0/0.0/15003/0000437. R.O.~acknowledges support from the COST Action GWverse CA16104.  
T.~H. and R.~O.~thank Perugia University and G.~G., R.~O.~and M.~O.~thank Niels Bohr Institute for hospitality.

\appendix
\section{Perturbative expressions of fields and constraints} \label{App:A}
The definition for the inverse of the metric $g^{\mu\nu}$ leads to
\begin{align}
\delta^\mu_\nu&=g^{\mu\alpha}g_{\alpha\nu} \nonumber\\
&=\big[(g^{\mu\alpha})^{(0)}+\lambda (g^{\mu\alpha})^{(1)}+\lambda^2(g^{\mu\alpha})^{(2)}+\dots\big]\big[(g_{\alpha\nu})^{(0)}+\lambda (g_{\alpha\nu})^{(1)}+\lambda^2(g_{\alpha\nu})^{(2)}+\dots\big] \nonumber\\
&=\delta^\mu_\nu+\lambda \big[(g^{\mu\alpha})^{(0)}(g_{\alpha\nu})^{(1)}+(g^{\mu\alpha})^{(1)}(g_{\alpha\nu})^{(0)}\big]\nonumber\\
&\quad\quad +\lambda^2 \big[(g^{\mu\alpha})^{(0)}(g_{\alpha\nu})^{(2)}+(g^{\mu\alpha})^{(2)}(g_{\alpha\nu})^{(0)}+(g^{\mu\alpha})^{(1)}(g_{\alpha\nu})^{(1)}\big]+ \mathcal{O}(\lambda^3),
\end{align}
that implies that the 1st and 2nd post-NHEK orders of the metric and its inverse obey the following constraints
\begin{subequations}
\begin{align}
(g^{\mu\alpha})^{(0)}(g_{\alpha\nu})^{(1)}+(g^{\mu\alpha})^{(1)}(g_{\alpha\nu})^{(0)}=0,\\
(g^{\mu\alpha})^{(0)}(g_{\alpha\nu})^{(2)}+(g^{\mu\alpha})^{(2)}(g_{\alpha\nu})^{(0)}+(g^{\mu\alpha})^{(1)}(g_{\alpha\nu})^{(1)}=0,
\end{align}
\end{subequations}
and so on for higher orders in the $\lambda$ expansion.

As already noted, the expansion for the field behaves as
\begin{align}
F_{\mu\nu}=\lambda^{-1}\big(F_{\mu\nu}\big)^{(-1)}+\lambda^0\big(F_{\mu\nu}\big)^{(0)}+\lambda \big(F_{\mu\nu}\big)^{(1)}+ \mathcal{O}(\lambda^2).
\end{align}
Raising-up the indices, one gets
\begin{align}
F^{\mu\nu}&=g^{\mu\alpha}F_{\alpha\beta}g^{\beta\nu}\nonumber\\
&=\big[(g^{\mu\alpha})^{(0)}+\lambda (g^{\mu\alpha})^{(1)}+\lambda^2 (g^{\mu\alpha})^{(2)}+\dots\big]\times\nonumber\\
&\quad \times \left[\lambda^{-1}\big(F_{\alpha\beta}\big)^{(-1)}+\lambda^0\big(F_{\alpha\beta}\big)^{(0)}+\lambda\big(F_{\alpha\beta}\big)^{(1)}+\dots\right]\big[(g^{\beta\nu})^{(0)}+\lambda (g^{\beta\nu})^{(1)}+\lambda^2 (g^{\beta\nu})^{(2)}+\dots\big] \nonumber\\
&=\lambda^{-1}\Big[(g^{\alpha[\mu})^{(0)}(g^{\nu]\beta})^{(0)}\big(F_{\alpha\beta}\big)^{(-1)}\Big]\nonumber\\
&\quad +\lambda^{(0)}\Big[(g^{\alpha[\mu})^{(0)}(g^{\nu]\beta})^{(0)}\big(F_{\alpha\beta}\big)^{(0)}+2(g^{\alpha[\mu})^{(1)}(g^{\nu]\beta})^{(0)}\big(F_{\alpha\beta}\big)^{(-1)}\Big] \nonumber\\
&\quad +\lambda\left[(g^{\alpha[\mu})^{(0)}(g^{\nu]\beta})^{(0)}\big(F_{\alpha\beta}\big)^{(1)}+2(g^{\alpha[\mu})^{(1)}(g^{\nu]\beta})^{(0)}\big(F_{\alpha\beta}\big)^{(0)}+2(g^{\alpha[\mu})^{(2)}(g^{\nu]\beta})^{(0)}\big(F_{\alpha\beta}\big)^{(-1)}\right.\nonumber\\
&\quad\left.+(g^{\alpha[\mu})^{(1)}(g^{\nu]\beta})^{(1)}\big(F_{\alpha\beta}\big)^{(-1)} \right] +\mathcal{O}(\lambda^2),
\end{align}
and we define, respectively, the NHEK, 1st and 2nd post-NHEK orders as
\begin{subequations} \label{FUUexpansion}
\begin{align}
\big(F^{\mu\nu}\big)^{(-1)} &= (g^{\alpha[\mu})^{(0)}(g^{\nu]\beta})^{(0)}\big(F_{\alpha\beta}\big)^{(-1)},\\
\big(F^{\mu\nu}\big)^{(0)} &= (g^{\alpha[\mu})^{(0)}(g^{\nu]\beta})^{(0)}\big(F_{\alpha\beta}\big)^{(0)}+2(g^{\alpha[\mu})^{(1)}(g^{\nu]\beta})^{(0)}\big(F_{\alpha\beta}\big)^{(-1)},\\
\big(F^{\mu\nu}\big)^{(1)} &=(g^{\alpha[\mu})^{(0)}(g^{\nu]\beta})^{(0)}\big(F_{\alpha\beta}\big)^{(1)}+2(g^{\alpha[\mu})^{(1)}(g^{\nu]\beta})^{(0)}\big(F_{\alpha\beta}\big)^{(0)}+\nonumber\\
&\quad +2(g^{\alpha[\mu})^{(2)}(g^{\nu]\beta})^{(0)}\big(F_{\alpha\beta}\big)^{(-1)}+(g^{\alpha[\mu})^{(1)}(g^{\nu]\beta})^{(1)}\big(F_{\alpha\beta}\big)^{(-1)}.
\end{align}
\end{subequations}

The way in which the metric and its inverse transform also affects covariant derivatives; for example, for what concerns the current 
\begin{align}
j^\mu&=\Big[\big(D_\nu\big)^{(0)}+\lambda\big(D_\nu\big)^{(1)}+\lambda^2\big(D_\nu\big)^{(2)}\dots\Big] \Big[\lambda^{-1}\big(F^{\nu\mu}\big)^{(-1)}+\lambda^0\big(F^{\nu\mu}\big)^{(0)}+\lambda\big(F^{\nu\mu}\big)^{(1)}+\dots\Big]
\nonumber\\
&=\lambda^{-1}\Big[\big(D_\nu\big)^{(0)}\big(F^{\nu\mu}\big)^{(-1)}\Big]
\nonumber\\
&\quad +\lambda^0\Big[\big(D_\nu\big)^{(0)}\big(F^{\nu\mu}\big)^{(0)}+\big(D_\nu\big)^{(1)}\big(F^{\nu\mu}\big)^{(-1)}\Big]
\nonumber\\
&\quad +\lambda\Big[\big(D_\nu\big)^{(0)}\big(F^{\nu\mu}\big)^{(1)}+\big(D_\nu\big)^{(1)}\big(F^{\nu\mu}\big)^{(0)}+\big(D_\nu\big)^{(2)}\big(F^{\nu\mu}\big)^{(-1)}\Big] +\mathcal{O}(\lambda^2),
\end{align}
and we define, respectively, the NHEK, 1st and 2nd post-NHEK orders as
\begin{subequations}
\begin{align}
(j^{\mu})^{(-1)} &=\big(D_\nu\big)^{(0)}\big(F^{\nu\mu}\big)^{(-1)},
\\
(j^{\mu})^{(0)} &=\big(D_\nu\big)^{(0)}\big(F^{\nu\mu}\big)^{(0)}+\big(D_\nu\big)^{(1)}\big(F^{\nu\mu}\big)^{(-1)},
\\
(j^{\mu})^{(1)} &=\big(D_\nu\big)^{(0)}\big(F^{\nu\mu}\big)^{(1)}+\big(D_\nu\big)^{(1)}\big(F^{\nu\mu}\big)^{(0)}+\big(D_\nu\big)^{(2)}\big(F^{\nu\mu}\big)^{(-1)},
\end{align}
\end{subequations}
where $\big(D_\nu\big)^{(n)}$ for $n>0$ stand for the expansion for the Christoffel symbols.

The FF condition implies that
\begin{align} \label{FFexp}
0=F_{\mu\nu}j^\nu&=\Big[\lambda^{-1}\big(F_{\mu\nu}\big)^{(-1)}+\lambda^0\big(F_{\mu\nu}\big)^{(0)}+\lambda\big(F_{\mu\nu}\big)^{(1)}+\dots\Big]\times\nonumber\\
&\qquad \times\Big[\lambda^{(-1)}\big(j^\nu\big)^{(-1)}+\lambda^0 \big(j^\nu\big)^{(0)}+\lambda \big(j^\nu\big)^{(1)} + \dots \Big] \\
&=\lambda^{-2}\left[\big(F_{\mu\nu}\big)^{(-1)}\big(j^\nu\big)^{(-1)}\right]\nonumber\\
&\quad +\lambda^{-1}\Big[\big(F_{\mu\nu}\big)^{(-1)}\big(j^\nu\big)^{(0)}+\big(F_{\mu\nu}\big)^{(0)}\big(j^\nu\big)^{(-1)}\Big]\nonumber\\
&\quad +\lambda^{0}\Big[\big(F_{\mu\nu}\big)^{(-1)}\big(j^\nu\big)^{(1)}+\big(F_{\mu\nu}\big)^{(0)}\big(j^\nu\big)^{(0)}+\big(F_{\mu\nu}\big)^{(1)}\big(j^\nu\big)^{(-1)}\Big]\nonumber\\
&\quad + \mathcal{O}(\lambda),
\end{align}
and we define
\begin{subequations}
\begin{align}
(F_{\mu\nu}j^\nu)^{(-2)}&=\big(F_{\mu\nu}\big)^{(-1)}\big(j^\nu\big)^{(-1)},\\
(F_{\mu\nu}j^\nu)^{(-1)}&=\big(F_{\mu\nu}\big)^{(-1)}\big(j^\nu\big)^{(0)}+\big(F_{\mu\nu}\big)^{(0)}\big(j^\nu\big)^{(-1)},\\
(F_{\mu\nu}j^\nu)^{(0)}&=\big(F_{\mu\nu}\big)^{(-1)}\big(j^\nu\big)^{(1)}+\big(F_{\mu\nu}\big)^{(0)}\big(j^\nu\big)^{(0)}+\big(F_{\mu\nu}\big)^{(1)}\big(j^\nu\big)^{(-1)}.
\end{align}
\end{subequations}

The Lorentz invariant $F^2$ is then given by
\begin{align} \label{F2exp}
    F_{\mu\nu}F^{\mu\nu} &= \lambda^{-2}\big(F_{\mu\nu}\big)^{(-1)}\big(F^{\mu\nu}\big)^{(-1)} + \lambda^{-1}\left[\big(F_{\mu\nu}\big)^{(-1)}\big(F^{\mu\nu}\big)^{(0)} + \big(F_{\mu\nu}\big)^{(0)}\big(F^{\mu\nu}\big)^{(-1)}\right]\nonumber\\
    &\quad +\lambda^{0}\left[\big(F_{\mu\nu}\big)^{(-1)}\big(F^{\mu\nu}\big)^{(-1)}+\big(F_{\mu\nu}\big)^{(0)}\big(F^{\mu\nu}\big)^{(0)}+\big(F_{\mu\nu}\big)^{(1)}\big(F^{\mu\nu}\big)^{(-1)}\right],
\end{align}
and we define
\begin{subequations}
\begin{align}
    \big(F^2\big)^{(-2)} &=\big(F_{\mu\nu}\big)^{(-1)}\big(F^{\mu\nu}\big)^{(-1)},\\
    \big(F^2\big)^{(-1)} &=\big(F_{\mu\nu}\big)^{(-1)}\big(F^{\mu\nu}\big)^{(0)} + \big(F_{\mu\nu}\big)^{(0)}\big(F^{\mu\nu}\big)^{(-1)},\\
    \big(F^2\big)^{(0)} &=\big(F_{\mu\nu}\big)^{(-1)}\big(F^{\mu\nu}\big)^{(1)}+\big(F_{\mu\nu}\big)^{(0)}\big(F^{\mu\nu}\big)^{(0)}+\big(F_{\mu\nu}\big)^{(1)}\big(F^{\mu\nu}\big)^{(-1)},
\end{align}
\end{subequations}
where the coefficients of the expansion of $F^{\mu\nu}$ (they involve the metric field expansion as well) are listed in Eqs.~\eqref{FUUexpansion}.

\section{The case with $\psi_0$ constant} \label{App:B}
Here we consider the case defined by the condition $\psi_0'=0$. The post-NHEK procedure, as outlined in Sec.~\ref{Sec:Climbing}, applies in the same fashion.
The main feature of this case is that the equations of motion of the $n$-th post-NHEK order unambiguously determine the field variables $(\psi_{n-1},I_{n-1},\Omega_{n-1})$ of the $(n-1)$th post-NHEK order. This contrasts with the case $\psi'_0 \neq 0$, where the field variables $(\psi_n,I_n,\Omega_n)$ of the $n$-th post-NHEK order are determined in terms of the unconstrained arbitrary NHEK functions $(\psi_0,\Omega_0)$.

Referring to the leading contribution in the expansion \eqref{expansions}, and assuming $\psi_0'=0$, the field strength \eqref{Fm1} and its associated current \eqref{jm1} are given by
\begin{equation}
    F^{(-1)}=r_0 I_0\frac{dR}{R^2}\wedge d\theta, \quad j^{(-1)}=-\frac{4 I_0'}{r_0^3\Gamma^2\Lambda}\partial_R.
\end{equation}
It is immediate to see that the FF condition amounts to $\left(F\cdot j\right)^{(-2)}\propto I_0 I_0'=0$, whose solution is $I_0=\iota_0$ constant. The Znajek condition \eqref{ZCExt} would imply $\iota_0=0$; however, at this stage, we leave $\iota_0$ unconstrained and we show that the regularity of the field strength on the horizon will naturally appear at the subsequent order when enforcing the FF condition.

At the next order in $\lambda$, the Bianchi identity reads $\left(dF\right)^{(0)}\propto \psi_1 \cdot r_0 \Omega_0'=0$, with the non-trivial solution given by $r_0\Omega_0=1-c_0$, and $c_0$ an arbitrary constant. 
The FF condition can be put in the compact form 
\begin{equation}
    \left(F\cdot j\right)^{(-1)}=\frac{2\iota_0}{r_0 R \Gamma^2 \Lambda^2}\left(I_1 \frac{dR}{R}+I_1' d\theta\right)=0.
\end{equation} 
The only solution consistent with regularity at the horizon, as imposed by Eq.~\eqref{ZCExt}, is $\iota_0=0$.
Thus, the equations of the 1st post-NHEK order fully determines the NHEK field variables
\begin{equation}
    \psi_0'=0,\quad I_0=0,\quad \Omega_0=\frac{1-c_0}{r_0},
\end{equation}
and lead to a vanishing NHEK field, \emph{i.e.}, $F^{(-1)}=0$.
The leading order contributions to the field and its associated current, therefore, come from
\begin{subequations}
\begin{align}
\label{1pn_F_B}
   & F^{(0)}=\frac{r_0 c_0}{2}\left(\psi_1dR+R\psi_1'd\theta\right)\wedge dT+\frac{I_1 r_0^2 }{2\Lambda R}dR\wedge d\theta,\\
   & j^{(0)}= -\frac{2}{r_0^2  \Gamma^2\Lambda}\left(c_0\frac{\Lambda'\psi_1'+\Lambda \psi_1''}{r_0 R}\partial_T+R I_1'\partial_R-I_1\partial_\theta-c_0\frac{\Lambda'\psi_1'+\Lambda\left(\psi_1+\psi_1''\right)}{r_0}\partial_\Phi\right),
\end{align}
\end{subequations}
with the field variables $(\psi_1,I_1,\Omega_1)$ that will be explicitly determined at the next post-NHEK order.

The Bianchi identity $\left(dF\right)^{(1)}=0$ relates linearly $\psi_1$ and $\Omega_1$ according to
\begin{equation}
\label{Omega1_B}
    \Omega_1=\frac{c_1}{r_0}\psi_1.
\end{equation}
As usual, from the components of the FF condition $\left(F\cdot j\right)^{(0)}=F^{(0)}\cdot j^{(0)}=0$ (recall Eq.~\eqref{FFexp} and the fact that $F^{(-1)} = 0 = j^{(-1)}$), one can extract the stream equation for $\psi_1$ and an integrability condition for $I_1$; solving these, respectively, yield to
\begin{subequations}
    \label{I1psi1_B}
    \begin{align}
    \label{psi1_B}
    \psi_1&=\frac{\kappa_1}{r_0}e^{-\iota_1\int \frac{d\theta}{\Lambda} },\\
    \label{I1_B}
    I_1&=\frac{c_0}{r_0}\iota_1\psi_1.
    \end{align}
\end{subequations}
It is worth to stress that this solution automatically satisfies regularity on the horizon, as expressed by Eq.~\eqref{ZCExt2}. 
Altogether Eqs.~\eqref{Omega1_B}-\eqref{I1psi1_B} serve us to write explicitly the leading order field and current vector as 
\begin{subequations}
    \begin{align}
        \label{newnhek}
         F^{(0)}&=-\frac{r_0 c_0}{2}\psi_1d\left(T-\frac{1}{R}\right)\wedge \left(dR-\iota_1 \frac{R}{\Lambda}d\theta\right),\\
   j^{(0)}&= \frac{2c_0 }{r_0^3 \Gamma^2\Lambda^2}\psi_1\left(-\frac{\iota_1^2}{R}\partial_T+\iota_1^2 R\partial_R+\iota_1\Lambda\partial_\theta +\left(\iota_1^2+\Lambda^2\right)\partial_\Phi\right).
    \end{align}
\end{subequations}
We remark that, to the best of our knowledge, this is a new solution to FFE in NHEK geometry.
It is readily shown from Eq.~\eqref{F2exp} (upon using $F^{(-1)} =0$) that this field strength is electrically-dominated
\begin{equation}
    \left(F^2\right)^{(0)}=-\frac{2 c_0^2}{r_0^2\Gamma^2}\psi_1^2<0.
\end{equation}
This feature motivated us not to consider the case $\psi_0'=0$ as relevant. Another physically motivated reason is that for $\psi'_0=0$ there is no extraction of energy and angular momentum from the horizon (see Eq.~\eqref{EJextraction}).

To conclude, we notice the special case $\iota_1=0$, for which Eq.~\eqref{1pn_F_B} simplifies to 
\begin{subequations}
\begin{align}
F^{(0)}&=-\frac{\kappa_1}{2}dT\wedge dR,\\
j^{(0)}&=\frac{2c_0}{r_0^3\Gamma^2}\partial_\Phi.
\end{align}
\end{subequations}
This is precisely the scale invariant field strength in Eq.~(23) of \cite{Gralla:2016jfc}, provided the identification of the constant $\kappa_1=4Q_E/\pi$ is made. 

\section{Expressions in the 2nd post-NHEK order} \label{Expr2ndNHEK}

The current vector at the 2nd post-NHEK expansion reads
\begin{equation}
    j^{(1)} = j_{(1)}^T\partial_T+R^2j_{(1)}^R\partial_R+Rj_{(1)}^\theta\partial_\theta+Rj_{(1)}^\Phi\partial_\Phi,
\end{equation}
where the explicit expressions of the components are
\begin{subequations}
\begin{align}
    j^T_{(1)}&=\partial_\theta\left[\frac{\Gamma(2+r_0^2\Omega_1)-4(1-\Gamma)(r_0\Omega_0-1)}{\Gamma^3}\psi_1'-\frac{4(1-\Gamma)r_0\psi_0'}{r_0^3\Gamma^3}\Omega_1\right] \nonumber \\
    \nonumber
    &+\frac{\Gamma(2+r_0\Omega_1)-2(1-2\Gamma)(r_0\Omega_0-1)}{r_0^3\Gamma^3}\left(\psi_1+2\frac{\Gamma'\psi_1'}{\Gamma}\right)\nonumber\\ \nonumber
    &+\frac{8\psi_0'}{r_0^4\Gamma^2}\left(\frac{\Lambda'}{\Lambda}+\frac{\psi_0''}{\psi_0'}+\frac{r_0I'_0}{\Lambda\psi_0'}\right)\left(1-\frac{1}{\Gamma}\right)+\frac{\partial_\theta\left(\Lambda\psi_0'\Omega_2\right)}{2r_0\Gamma^2\Lambda}
    \\
    \nonumber
    &+\frac{\psi_1'}{r_0^2\Gamma^2}\left[\frac{(2+r_0^2\Omega_1)\Lambda'}{r_0\Lambda}-\frac{2I_0}{\Gamma\Lambda\psi_0'}\left(3\frac{\Gamma'}{\Gamma}+2(1-\Gamma)\frac{\Lambda'}{\Lambda}\right)\right]-\frac{12}{r_0^3\Gamma^3\Lambda}\partial_\theta\left[I_0\left(1-\frac{1}{\Gamma}\right)\right]
    \\
    \nonumber
    &-\frac{2\Omega_1\psi_0'}{r_0^2\Gamma^3}\left[2(1-\Gamma)\frac{\Lambda'}{\Lambda}+(5-4\Gamma)\frac{\Gamma'}{\Gamma}\right]+\frac{I_0}{2r_0\Gamma^2\Lambda\psi'_0}\left[\psi_2''+\psi_2'\left(\frac{I_0'}{I_0}-\frac{\psi_0''}{\psi_0'}\right)+2\psi_2\right]
    \\
    &-\frac{2I_0\Lambda}{r_0^3}\left(\frac{\Lambda'}{\Lambda}+\frac{I_0'}{2I_0}+\frac{\Gamma'}{\Gamma}\right)+\frac{\Lambda^2\psi_0'}{r_0^4}\left(\frac{\psi_0''}{\psi_0'}+3\frac{\Lambda'}{\Lambda}\right)-\frac{2\Gamma'\psi_0'}{r_0^4\Gamma^4}(3+\Gamma^3\Lambda^2-5r_0\Omega_0),
    \\
    j_{(1)}^R&=\frac{r_0\Gamma(4I_1'-r_0\Gamma I_2')-4(2-\Gamma)I_0'}{2r_0^3\Gamma^4\Lambda},
    \\
    j_{(1)}^\theta&=\frac{r_0I_2\Gamma-2I_1}{r_0^2\Gamma^3\Lambda},
    \\
    j_{(1)}^\Phi&=- j_{(1)}^T+\frac{\psi_0'}{2r_0^4\Gamma^3\Lambda^2}\bigg\{-8\frac{\Gamma'}{\Gamma}+2\Gamma^2\Lambda^3\left[(2+r_0^2\Omega_1)\Lambda\Gamma'-2r_0\frac{I_0'}{\psi_0'}\right]
    \nonumber \\
    \nonumber
    &+\Gamma^3\Lambda^3\left[r_0\frac{I_0'}{\psi_0'}\left(r_0\frac{\psi_1'}{\psi_0'}-2\right) +r_0^2\Lambda\Omega_1'+\Lambda\left(3\frac{\Lambda'}{\Lambda}+\frac{\psi_0''}{\psi_0'}-\frac{2r_0}{\Gamma^2\Lambda^2}\frac{\psi_1}{\psi_0'}\right)(2+r_0^2\Omega_1)\right]
    \\
    \nonumber
    &+4\Gamma\left[2r_0\Gamma\frac{I_0'}{\psi_0'}+\frac{\Lambda'}{\Lambda}\left(4-r_0\frac{\psi_1'}{\psi_0'}\right)+\left(2r_0\frac{\psi_1}{\psi_0'}-4\frac{\psi_0''}{\psi_0'}+r_0\frac{\psi_1''}{\psi_0'}\right)\right]
    \\
    \nonumber
    &+2r_0\frac{I_0}{\psi_0'}\Gamma^3\Lambda^3\left[2\frac{(2+\Gamma)}{\Gamma}\frac{\Lambda'}{\Lambda}+\frac{(3+2\Gamma)}{\Gamma}\frac{\Gamma'}{\Gamma}\right]-r_0^2I_0\frac{\psi_1'}{\psi_0'^2}\Gamma^3\Lambda^3\left(\frac{\psi_0''}{\psi_0'}-2\frac{\Lambda'}{\Lambda}-2\frac{\Gamma'}{\Gamma}-\frac{\psi_1''}{\psi_1'}\right)
    \\
    &+2r_0I_0\frac{\psi_1}{\psi_0'^2}\Gamma^3\Lambda^3\left(1+\frac{2}{\Gamma}-\frac{r_0}{\Lambda^2\Gamma^2}\frac{\psi_2}{\psi_1}\right)\bigg\}.
\end{align}    
\end{subequations}
The expressions $A$, $B$, and $C$ present in the coefficients \eqref{psi2_coeffs} of the differential equation \eqref{psi2_eq} read as
\begin{subequations}
\begin{align}
    A(\theta) &= 8\left(\frac{\psi'_0}{I_0} \right)^2 \left(\frac{4\Lambda^2-1}{\Lambda^2}\frac{\Lambda'}{\Lambda} +\frac{4\Lambda^2+1}{\Lambda^2}\frac{\psi''_0}{\psi'_0} -\frac{I'_0}{I_0} \right)\nonumber\\
    &\quad +2r_0\Lambda \Gamma^2\frac{\psi'_0}{I_0}\left[4\left(1+\frac{1}{\Gamma^3\Lambda^2}\right)\frac{\Gamma'}{\Gamma} + 5\frac{\Lambda'}{\Lambda}+\frac{\psi''_0}{\psi'_0}+\frac{I'_0}{I_0} -\frac{8}{\Gamma}\left( \frac{\Lambda'}{\Lambda} + \frac{\psi''_0}{\psi'_0}\right) \right]\nonumber\\
    &\quad + 4r_0^2\Gamma^2 \left[\left(1+\frac{1}{\Gamma^3\Lambda^2}\right)\frac{\Gamma'}{\Gamma} + \frac{\Lambda'}{\Lambda}+\left(\frac{1}{2}-\frac{1}{\Gamma}\right)\frac{I'_0}{I_0}+\frac{\Lambda^2}{2}\left(\frac{\Lambda'}{\Lambda}+\frac{\psi''_0}{\psi'_0}\right) \right],\\
    B(\theta)&=-4r_0\frac{\psi'_0}{I_0}\left[\frac{\Lambda'}{\Lambda}\left(2\frac{\Lambda'}{\Lambda}+3\frac{\Lambda''}{\Lambda'}+7\frac{\psi''_0}{\psi'_0}\right)+2-\left( \frac{\psi''_0}{\psi'_0}\right)^2 +3\frac{\psi^{(3)}_0}{\psi'_0} +\frac{I'_0}{I_0}\left(\frac{\psi''_0}{\psi'_0}-\frac{I''_0}{I'_0} \right) \right]\nonumber\\
    &\quad +4r_0^2\Lambda \Gamma \left[1-\frac{1}{\Gamma^2 \Lambda^2}- \left(\frac{1}{\Gamma^2\Lambda^2}\frac{\Gamma'}{\Gamma} - \frac{I'_0}{I_0}\right)\left( \frac{\Lambda'}{\Lambda}+\frac{\psi''_0}{\psi'_0}\right) + \frac{\Lambda'}{\Lambda}\left(\frac{\Lambda''}{\Lambda'}+\frac{\psi''_0}{\psi'_0} \right) -\left(\frac{\psi''_0}{\psi'_0}\right)^2 + \frac{\psi^{(3)}}{\psi'_0} \right],\\
    C(\theta) &= r_0^2\Bigg\{\frac{\Lambda ''}{\Lambda } \left(\frac{\Lambda^{(3)}}{\Lambda ''}+\frac{\psi''_0}{\psi'_0}\right)+\frac{\Lambda'}{\Lambda} \left[\left(\frac{\Lambda '}{\Lambda }+2\frac{ I'_0}{I_0}\right) \left(\frac{\Lambda ''}{\Lambda '}+\frac{\psi''_0}{\psi'_0}\right)-5\left(\frac{\psi''_0}{\psi'_0}\right)^2+6\frac{\psi^{(3)}_0}{\psi'_0}\right]\nonumber\\
    &\quad \quad + \left[2+2\left(\frac{ \psi''_0}{\psi'_0}\right)^2-3\frac{ \psi_0^{(3)}}{\psi'_0}\right] \left(\frac{\Lambda '}{\Lambda }+\frac{\psi''_0}{\psi'_0}-\frac{I'_0}{I_0}\right)  - \frac{I'_0}{I_0}\frac{\psi^{(3)}_0}{\psi'_0} +\frac{\psi^{(4)}_0}{\psi'_0}\Bigg\}.
\end{align}
\end{subequations}
The coefficients $D$ and $E$ in Eq.~\eqref{F2} are explicitly given by
\begin{subequations}
\begin{align}
    D(\theta) &=4\left(1-\Lambda^2 \right)+2r_0(\Gamma+4)\Gamma \Lambda^3 \frac{I_0}{\psi'_0}+r_0^2\left(-\frac{12}{\Gamma^2}+\frac{20}{\Gamma}+\Gamma^2\Lambda^2-8 \right)\left(\frac{I_0}{\psi'_0}\right)^2 \nonumber\\
    &\quad + r_0^4\frac{\psi_2}{\psi'_0}\left(\frac{\Lambda'}{\Lambda}+\frac{\psi''_0}{\psi'_0}-\frac{\psi'_2}{\psi_2} \right) \left(\frac{I_0}{\psi'_0}\right)^2,\\
    E(\theta)&=r_0\Lambda^2\frac{I_0}{\psi'_0}\Bigg\{-4\left( \frac{\Lambda'}{\Lambda}+\frac{\psi''_0}{\psi'_0}+\frac{I'_0}{I_0}+2\frac{\mathcal{G'}}{\mathcal{G}}\right)\nonumber\\
    &\quad\qquad \qquad+2r_0\mathcal{G'}\frac{I_0}{\psi'_0}\left[\frac{\Lambda'}{\Lambda}+\frac{\psi''_0}{\psi'_0}-2\frac{I'_0}{I_0}+2\left(\frac{I'_0}{I_0}+\frac{\mathcal{G'}}{\mathcal{G}}\right)\frac{\Gamma \Lambda}{\mathcal{G'}}-\frac{1}{2} \frac{\mathcal{G'}}{\mathcal{G}} \right]\nonumber\\
    &\quad\qquad \qquad+r_0\mathcal{G}\frac{I_0}{\psi'_0}\left[\frac{\Lambda'}{\Lambda}\left(\frac{\Lambda''}{\Lambda'}+ \frac{\psi''_0}{\psi'_0}\right) -1-\left(\frac{\psi''_0}{\psi'_0}\right)^2+\frac{\psi^{(3)}_0}{\psi'_0} \right] \Bigg\}.
\end{align}
\end{subequations}

\newpage
\bibliography{Bibliography}

\end{document}